\documentclass[preprint,aps,showpacs,superscriptaddress]{revtex4-1}

\usepackage[english]{babel}
\usepackage[utf8x]{inputenc}
\usepackage[T1]{fontenc}
\usepackage{amsmath,amssymb}

\usepackage{relsize}
\usepackage{natbib}
\usepackage[a4paper, portrait, margin=0.5in]{geometry}

\usepackage{graphicx}
\usepackage[colorinlistoftodos]{todonotes}
\usepackage[colorlinks=true, allcolors=blue]{hyperref}
\usepackage{lipsum}
\usepackage{mathtools}
\usepackage{parskip}
\newcommand{\mem}{K}
\newcommand{\chim}{\chi_m}
\newcommand{\U}{U_s}
\newcommand{\A}{A}
\newcommand{\Sweep}{S}
\newcommand{\Burn}{S_b}

\begin{document}
\title{Kinetic Ising Models with Self-interaction: Sequential and Parallel Updating}

\author{Vahini Reddy Nareddy}
\email{vnareddy@umass.edu}
\affiliation{Department of Physics, University of Massachusetts,
Amherst, Massachusetts 01003 USA}
\author{Jonathan Machta}
\email{machta@physics.umass.edu}
\affiliation{Department of Physics, University of Massachusetts,
Amherst, Massachusetts 01003 USA}
\affiliation{Santa Fe Institute, 1399 Hyde Park Road, Santa Fe, New Mexico
87501, USA}
\begin{abstract}
Kinetic Ising models on the square lattice with both nearest-neighbor interactions and self-interaction are studied for the cases of random sequential updating and parallel updating. The equilibrium phase diagrams and critical dynamics are studied using Monte Carlo simulations and analytic approximations. The Hamiltonians appearing in the Gibbs distribution describing the equilibrium properties differs for sequential and parallel updating but in both cases feature multispin and non-nearest-neighbor couplings.  For parallel updating the system is a probabilistic cellular automaton and the equilibrium distribution satisfies detailed balance with respect to the dynamics [E. N. M. Cirillo, P. Y. Louis, W. M. Ruszel and C. Spitoni, Chaos, Solitons and Fractals, 64:36(2014)].  In the limit of weak self-interaction for parallel dynamics, odd and even sublattices are nearly decoupled and checkerboard patterns are present in the critical and low temperature regimes, leading to singular behavior in the shape of the critical line.  For sequential updating the equilibrium Gibbs distribution satisfies global balance but not detailed balance and the Hamiltonian is obtained perturbatively in the limit of weak nearest-neighbor dynamical interactions.  In the limit of strong self-interaction the equilibrium properties for both parallel and sequential updating are described by a nearest-neighbor Hamiltonian with twice the interaction strength of the dynamical model. 

\end{abstract}


\maketitle
\section{Introduction}

Kinetic Ising models, broadly defined, are systems of Ising spins equipped with a dynamical rule for updating the spin configuration.  Kinetic Ising models have been investigated in many settings and play a fundamental role in understanding non-equilibrium processes, in modelling the dynamics of more complex systems and as computational tools for studying equilibrium systems through the use of Markov chain Monte Carlo.  The most commonly used kinetic Ising models update spins sequentially and determine the new spin state probabilistically from the current state of the neighboring spins using update rules such as heat bath dynamics or Metropolis dynamics.  In this paper we study kinetic Ising models with a self-interaction or memory term so that the new state of the spin also depends on the current state of the spin itself.  We consider this kinetic Ising model both with random sequential and parallel updating.  The parallel version of the model is an example of a probabilistic cellular automata and many of its properties are understood \citep{Jayaprakash,CIRILLO201436}.  The sequential version of the model violates detailed balance when the self-interaction term is finite but non-vanishing and has received less attention.

Dynamical Ising-like systems with self-interaction have been proposed to model a variety of systems including neural networks \citep{StSoAb14}, financial markets \citep{Born2001} and opinion dynamics \cite{Caccioli_2008}.  The addition of a memory term has also be proposed as a method to accelerate Monte Carlo simulations~\citep{momentumMC}. We were motivated to study the kinetic Ising model with memory as a reduced description of synchronous ecological populations described by coupled, noisy oscillators.  When each oscillator is individually in a two-cycle regime, the phase of the two-cycle can be represented by an Ising spin.   As a function of noise and coupling strength, these systems undergo a phase transition to synchronous oscillation, which can be mapped to an ordering transition in the Ising universality class \cite{Noble2015}.  However, to  model dynamic and non-universal properties of the coupled oscillator system it is necessary to include memory in the Ising representation since each oscillator remembers its own phase for an extended period, even in the absence of coupling to its neighbors.

The paper is organized as follows.  In Sec.\ \ref{sec:modelmethods}, we introduce the sequential and parallel kinetic Ising models with self-interaction and describe our numerical methods.  In Sec.\ \ref{Numericalresults} we present  numerical results for the equilibrium phase diagram of both models and the dynamic properties along the critical lines. In Sec.\ \ref{sec:equilibrium} we analyze the equilibrium states of the parallel and sequential models and derive associated Hamiltonians.  In the case of parallel dynamics we focus on the weak self-interaction regime where the critical line displays singular behavior.  For the sequential case, we derive the equilibrium  Hamiltonian from the global balance equations in a weak coupling expansion.  We also study the nontrivial case of parallel updating in standard Ising model with no memory. In Sec.\ \ref{sec:mft} we develop a mean field approximation to the kinetic Ising model with self-interaction. The paper concludes with a discussion.

\section{Model and Numerical Methods}
\label{sec:modelmethods}

 We consider kinetic Ising models with both nearest-neighbor interactions and self-interaction or, equivalently, memory.  The system consists of $N$ Ising spins,  $S_i = \pm 1$ and $i=1,2, \ldots, N$, on a two-dimensional square lattice of even size with periodic boundary conditions.
The dynamics of the system is defined by the transition probability, $\A(\alpha \rightarrow \gamma)$ between successive configurations of the system $\alpha$ and $\gamma$.  We consider two types of heat bath dynamics, parallel and sequential.  In both cases the transition probabilities are controlled by the single-spin marginal probability $p(S_i^\gamma | \alpha)$ for finding spin at site $i$ in the new configuration $\gamma$, $S_i^\gamma$,  given the initial configuration $\alpha$,
\begin{equation}
\label{eq:heatbath}
p(S_i^\gamma | \alpha) =  \frac{\exp[(J h_i^\alpha+\mem S_i^\alpha) S_i^\gamma]}{2\cosh(J h_i^\alpha+\mem S_i^\alpha)} .
\end{equation}
Here $J$ is the dimensionless nearest-neighbor interaction and $\mem$ is the dimensionless self-interaction.  The local field, $h_i^\alpha$ is the sum over spins $S_j^\alpha$ where $j$ is a neighboring site of $i$,
 \begin{equation}
h_i^\alpha = \sum_{\langle j;i \rangle 1} S_j^\alpha.
\end{equation}
The notation $\langle j;i \rangle 1$ indicates that the summation variable $j$ is a nearest neighbor of $i$.
In what follows we omit the superscripts specifying the configuration where no confusion results.

\paragraph*{Sequential dynamics}
For random sequential dynamics the transition probability $\A_s(\alpha \rightarrow \gamma)$ for $\gamma \neq \alpha$ is given by,
\begin{equation}\label{acceptanceratio}
\A_s(\alpha \rightarrow \gamma) = \frac{1}{N} \sum_i\chi_1(\alpha,\gamma,i)p(S_i^\gamma | \alpha) ,
\end{equation}
where $\chi_1(\alpha,\gamma,i)$ is an indicator function that forces $\alpha$ and $\gamma$ to differ only on $S_i$, that is $\chi_1(\alpha,\gamma,i)=1$ if $S_i^\gamma = - S_i^\alpha$ and $S_j^\gamma =  S_j^\alpha$ for all $j \neq i$, while $\chi_1(\alpha,\gamma,i)=0$ otherwise. The probability for no transition, $\A_s(\alpha \rightarrow \alpha)$ is obtained, as usual, from normalization. For $\mem=0$, sequential dynamics is standard heat bath dynamics.

In simulations, a single step of the sequential dynamics consists of choosing a site $i$ at random and then choosing the spin state $S_i$ according to the probability $p(S_i | \alpha)$. The result is the new state $\gamma$. 

\paragraph*{Parallel dynamics}
For parallel dynamics all spins are updated at the same time and the transition probability $\A_p(\alpha \rightarrow \gamma)$ is given by  
\begin{equation}
\label{eq:paralleldynamics}
\A_p(\alpha \rightarrow \gamma) = \prod_{i} p(S_i^\gamma | \alpha). 
\end{equation}
%In simulations, a single step of the parallel dynamics consists of updating each spin in the lattice in some order, setting the spin at $i$ according to the probabilities $p(S_i^\gamma | \alpha)$.  In this process $\alpha$ is unchanged until all spins have been updated.  
For $\mem=0$, parallel dynamics on a bipartite graph permits period-two oscillatory states.  For example, if $J$ is sufficiently large, the system may oscillate between the two ``checkerboard'' states (i.e.\ ground states of the Ising antiferromagnet).  The Ising model with parallel dynamics is an example of a probabilistic cellular automata \citep{CIRILLO201436,Cirillo_parallel_detail_balance,louis_thesis}.

We carried out extensive simulations of both parallel and sequential dynamics with the primary goal of understanding the static and dynamic critical properties of the models. 
To identify the critical points of the models for various values of $\mem$ we used the Binder cumulant method.  The fourth-order Binder cumulant for the magnetization is given by \citep{binder},
\begin{equation} \label{binder}
U = 1-\frac{\langle  M^4 \rangle}{3 \langle M^2 \rangle ^2},
\end{equation}
where $M = \frac{1}{N}\sum_i S_i$ is the magnetization per spin. At low temperatures, in the ferromagnetic phase, the Binder cumulant takes the value $2/3$ whereas it goes to zero in paramagnetic phase. At the phase transition, the critical Binder cumulant for the standard Ising model on the 2D square lattice with periodic boundary conditions is $U^* \approx 0.61069$ \citep{Selke2006}. The value of the critical coupling is obtained from crossing of the Binder cumulant curves for different system sizes. For parallel dynamics and small values of $\mem$ we also make use of a sublattice Binder cumulant to find critical points for reasons discussed in Sec.\ \ref{zero case}.

The dynamics of the sequential model are studied using the magnetization integrated autocorrelation time defined from the magnetization autocorrelation function,  $\Gamma_M(t)$.  Here time is measured in Monte Carlo sweeps and the integrated autocorrelation time, $\tau$ is defined as,
\begin{equation}
    \tau = \frac{1}{2}+ \sum_{t=1}^\infty \Gamma_M(t).
\end{equation}
Care must be taken in estimating $\tau$ from numerical data for $\Gamma_M(t)$ since an upper cut-off is required on the sum.  If the cut-off is too small the estimate will have large systematic errors and if it is too large it will have large statistical errors.  We follow the procedure described in Ref.\ \cite{OsSo04} for choosing the cut-off.

We have carried out simulations of both the parallel and sequential models for multiple values of $K > 0$ for lattice sizes varying from $L = 10$ to $120$. The critical dynamic couplings, $J_c^{(p)}(K)$ and $J_c^{(s)}(K)$, for the parallel and sequential cases, respectively, are estimated from the crossing points of the Binder cumulant curves for two lattice sizes. The crossing point were found for sizes $L=30$ and $60$, except for the data presented in Sec.\ \ref{zero case}, where two sizes were $L=100$ and 120. Each Binder cumulant data point is averaged over 40 independent runs for the sequential results and 20 independent runs for parallel results.  Each run consists $\Sweep$ Monte Carlo sweeps including an initial $\Burn$ sweeps for equilibration. Observables are averaged over the remaining $\Sweep-\Burn$ sweeps. For the sequential model, $\Sweep = 7 \times 10^7$ and $\Burn = 5 \times 10^7$ while, for the parallel model,  $\Sweep = 4 \times 10^7$ and $\Burn = 2 \times 10^7$ except for the data presented in Sec.\ \ref{zero case}
where  $\Sweep=8 \times 10^7$ and $\Burn = 5 \times 10^7$. 

\section{Static and Dynamic Critical Behavior}\label{Numericalresults}
% The aim of the work is to obtain the phase diagram with self-interaction $K$ separating ordered and disordered magnetic phases with sequential and parallel updating. Spins on two dimensional square lattice with periodic boundary conditions are updated with Markov Chain Monte Carlo methods using heat-bath algorithm. Autocorrelation time of magnetization at critical temperature due to self-interaction is also discussed here. The critical temperature $T_c$ for phase transition from ordered ferromagnetic to disordered paramagnetic phase is found using the fourth-order Binder cumulant given by \citep{binder}
% \begin{equation} \label{binder}
% U = 1-\frac{\langle  M^4 \rangle}{3 \langle M^2 \rangle ^2}
% \end{equation}
% where $M$ is the magnetization per spin on the lattice. At low temperatures, in ferromagnetic phase, the Binder cumulant takes the value $\frac{2}{3}$ where as it takes value close to zero in paramagnetic phase. At the phase transition, the critical Binder cumulant for the standard Ising model on 2D square lattice with periodic boundary conditions is $U^* = 0.61069$ \citep{Selke2006}.
% The critical temperature and critical Binder cumulant have finite size effects but the exact analysis is out of scope of this work.

\subsection{Phase diagrams}

\begin{figure}
\centering
\includegraphics[width = .5\textwidth]{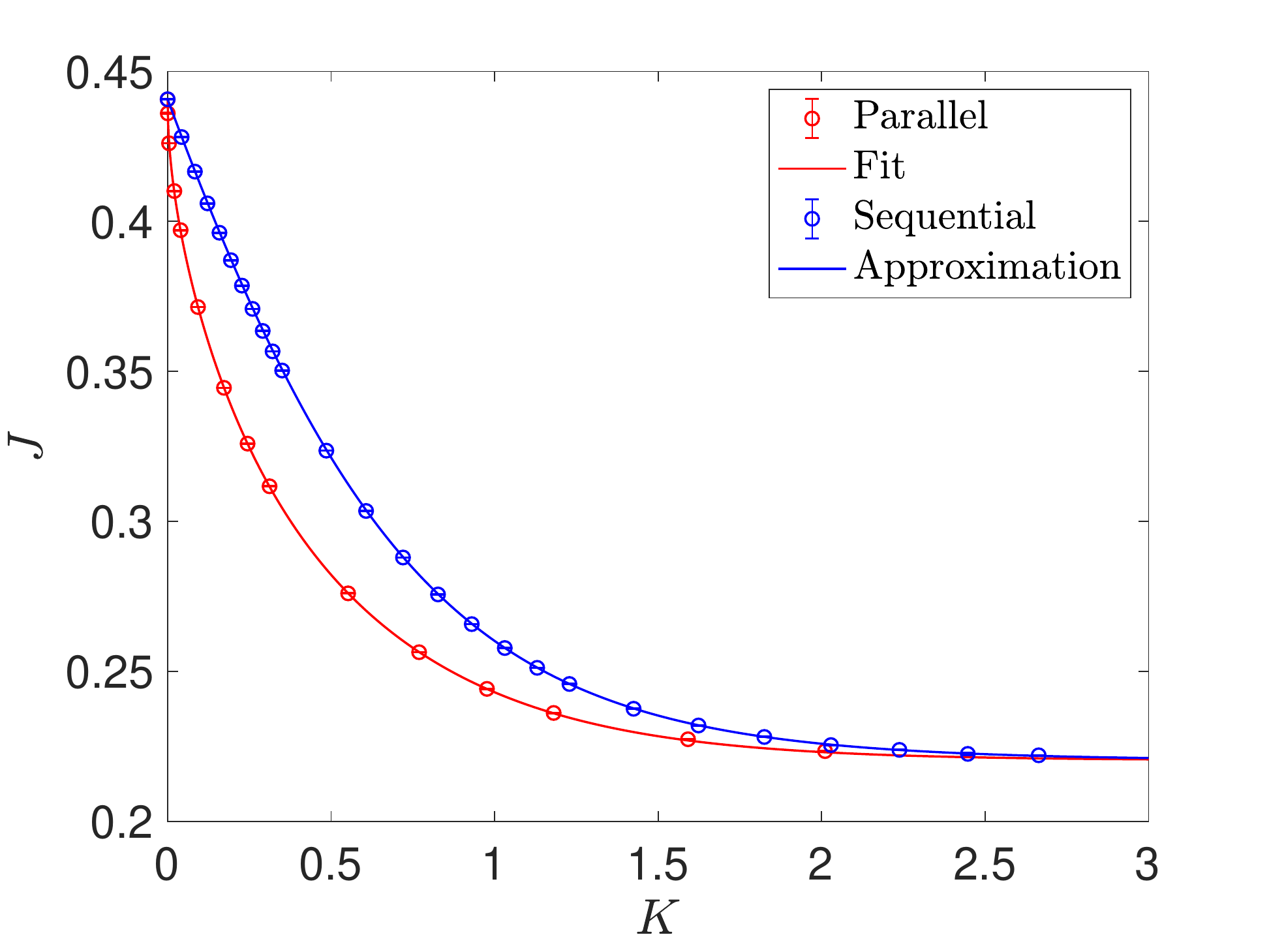}
\caption{\label{fig:phase_plot} The phase diagram in the $J$-$K$ plane shows the critical dynamical couplings, $J_c^{(s)}(K)$ and $J_c^{(p)}(K)$,  for sequential (blue points,upper) and parallel (red points,lower) dynamics, respectively. 
The red solid line represents a fit for the critical line for parallel dynamics (see Eq.\ \eqref{par_fit}). The blue solid line represents an approximation to the sequential  critical line that is described in Sec.\ \ref{sec:seqapprox}. 
}
\end{figure}

The phase diagrams for sequential and parallel dynamics are plotted in Fig.\ \ref{fig:phase_plot}. Critical values of the dynamic couplings, $J_c^{(s)}(\mem)$ and $J_c^{(p)}(\mem)$, for sequential and parallel dynamics, respectively, and for various values of $\mem$ are obtained from crossings of the Binder cumulant as discussed above.  The ferromagnetic phase lies above the points and the paramagnetic phase below the points.  As expected,  as $\mem \rightarrow 0$, the critical lines both approach $J_c(0) \approx 0.4407$, the critical value of the standard nearest-neighbor square-lattice Ising model.  Furthermore, as discussed in Sec.\ \ref{sec:equilibrium}, for $\mem$ large, both lines approach $J_c(0)/2$. The numerical result $J_c^{(p)}(\mem=0.31) \approx 0.31$ agrees well with the value  $J_c^{(p)}(K=0.31) \approx 0.32$ obtained from Monte Carlo simulations in Ref.\ \citep{louis_thesis}. 

The solid curves in Fig.\ \ref{fig:phase_plot} are approximations to the critical lines for the two dynamics.  For parallel dynamics, the approximation is a fit with two hyperbolic tangent functions that represent crossovers from power-law behavior expected for small $K$ and the asymptotic values $J_c(0)/2$ for large $K$.  The fit takes the form,
\begin{equation} \label{par_fit}
    J_c^{(p)}(K) = J_c(0) \big[1- \frac{1}{2} \tanh(a K)\big]-d K^{c} \big[1- \tanh(b K)\big],
\end{equation}
with the best fit parameters $a=1.17$, $b=1.46$, $c=0.45$, and $d=0.15$. A more detailed analysis of the power-law behavior in $\mem$ is presented in Sec.\ \ref{zero case}.

The solid curve in Fig.\ \ref{fig:phase_plot} for sequential dynamics is obtained from an approximation to the Hamiltonian for sequential dynamics that is obtained in Sec.\ \ref{sec:seqapprox} and is defined in Eq.\ \eqref{eq:seqJc}.
% Figure \ref{fig:phase_plot} shows the phase diagram separating ferromagnetic and paramagnetic phases for sequential and parallel updating. The results satisfy with the standard Ising model $\beta_c$ when the self-interaction is zero. $J$ decreases (or critical temperature increases) as self-interaction $K$ increases which is expected from the definition of self-interaction $K$. Note that $\beta_c$ for high values of $K$ drops to half its value at $K=0$ for both sequential and parallel updating. This behavior can be explained with the Hamiltonian couplings which were discussed in the Sections \ref{parallel} and \ref{sequential}.  It is shown in Figs.\ \ref{fig:parallel} and \ref{fig:seq_couplings} that the couplings saturate for higher K values and the nearest neighbor coupling increases to twice the value of J which is responsible for increasing/decreasing $T_c$/$\beta_c$ to twice/half the value of standard Ising $T_c$/$\beta_c$. The red solid line in Fig.\ \ref{fig:phase_plot} is the fit for the parallel numerical data with the expression,
% \begin{equation} \label{par_fit}
%     J = -0.15 K^{0.45} [1- \tanh{1.46 K}] + J_0 [1- 0.5 \tanh{1.17 K}]
% \end{equation}
% where $J_0 = 0.4407$ is the nearest neighbor critical coupling for standard Ising model.

We have not investigated the static critical properties along the critical lines however we find that the Binder cumulant takes the universal Ising value (except near $\mem=0$ for the parallel case, as discussed in Sec.\ \ref{zero case}), supporting general arguments \citep{Jayaprakash} that kinetic models with self-interaction are in the Ising universality class.

\subsection{Autocorrelation time for sequential dynamics}

Along the critical line for lattice size $L$, $\tau$ is expected to behave asymptotically in $L$ as a power law,
\begin{equation}\label{eqntau0}
    \tau(K,L) \sim A(K) L^{z(K)}
\end{equation}
where the amplitude $A$ and dynamic exponent $z$ could, in principle, both be functions of $K$. 

\begin{figure}
\centering
\includegraphics[width = .5\textwidth]{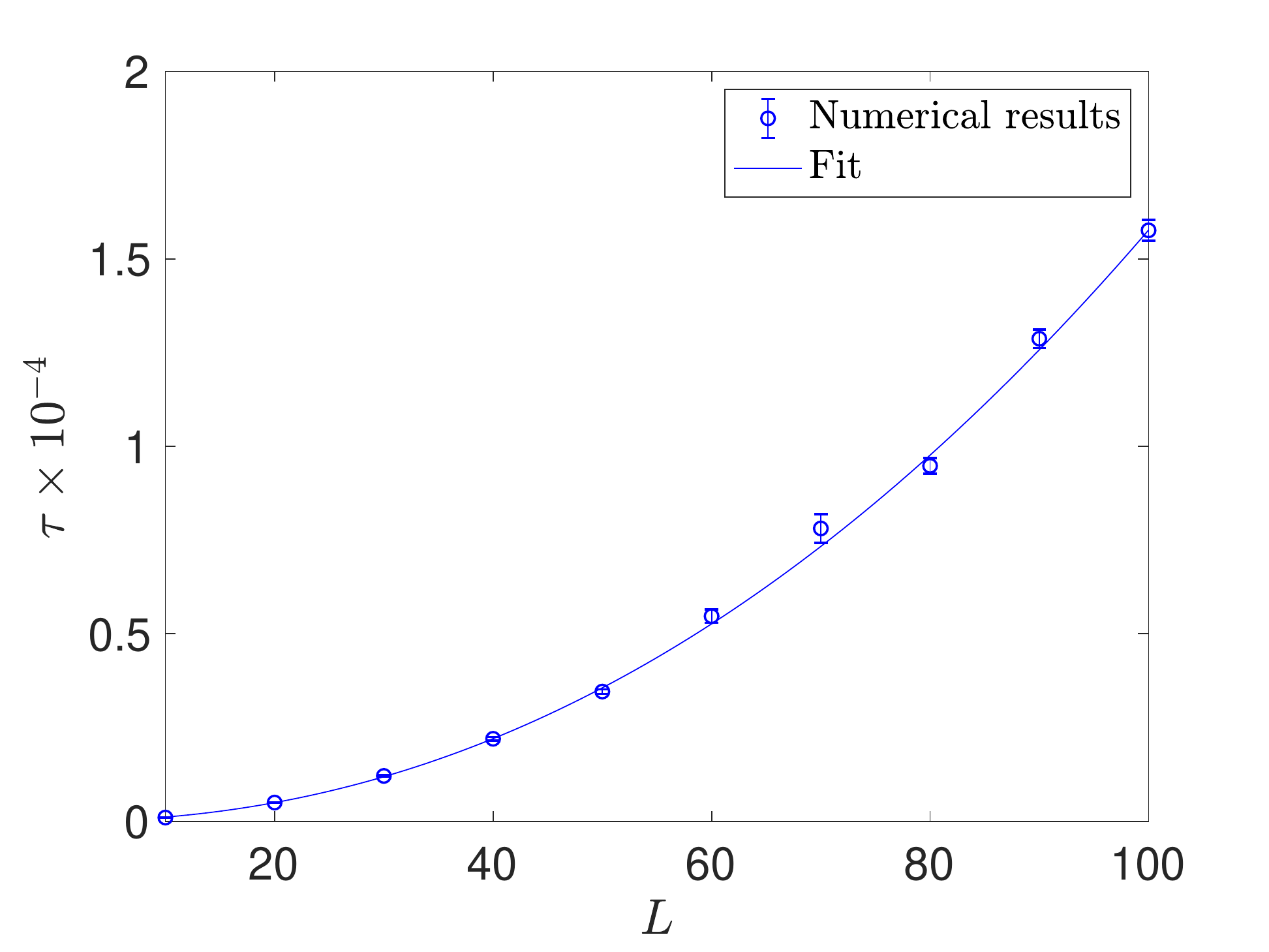}
\caption{\label{fig:zfit} The magnetization integrated autocorrelation time for sequential dynamics, $\tau$ as function of lattice size $L$ for $K=1.23$. The solid line is a fit to $\tau = A L^{z}$ with fitted values $A=0.16$ and $z=2.15$.}
\end{figure}

We obtained values of $\tau$ for lattice sizes from $L=10$ to 100 and for different values of $K$. The results for $\tau$ as a function of $L$ for $K=1.23$ are shown in Fig.\ \ref{fig:zfit}. The solid line is a fit to $\tau = A L^{z}$ with fitted values $z=2.15$ and $A=0.16$ ($\chi^2 /$ d.o.f = 1.38).

\begin{figure}
\centering
\includegraphics[width = .5\textwidth]{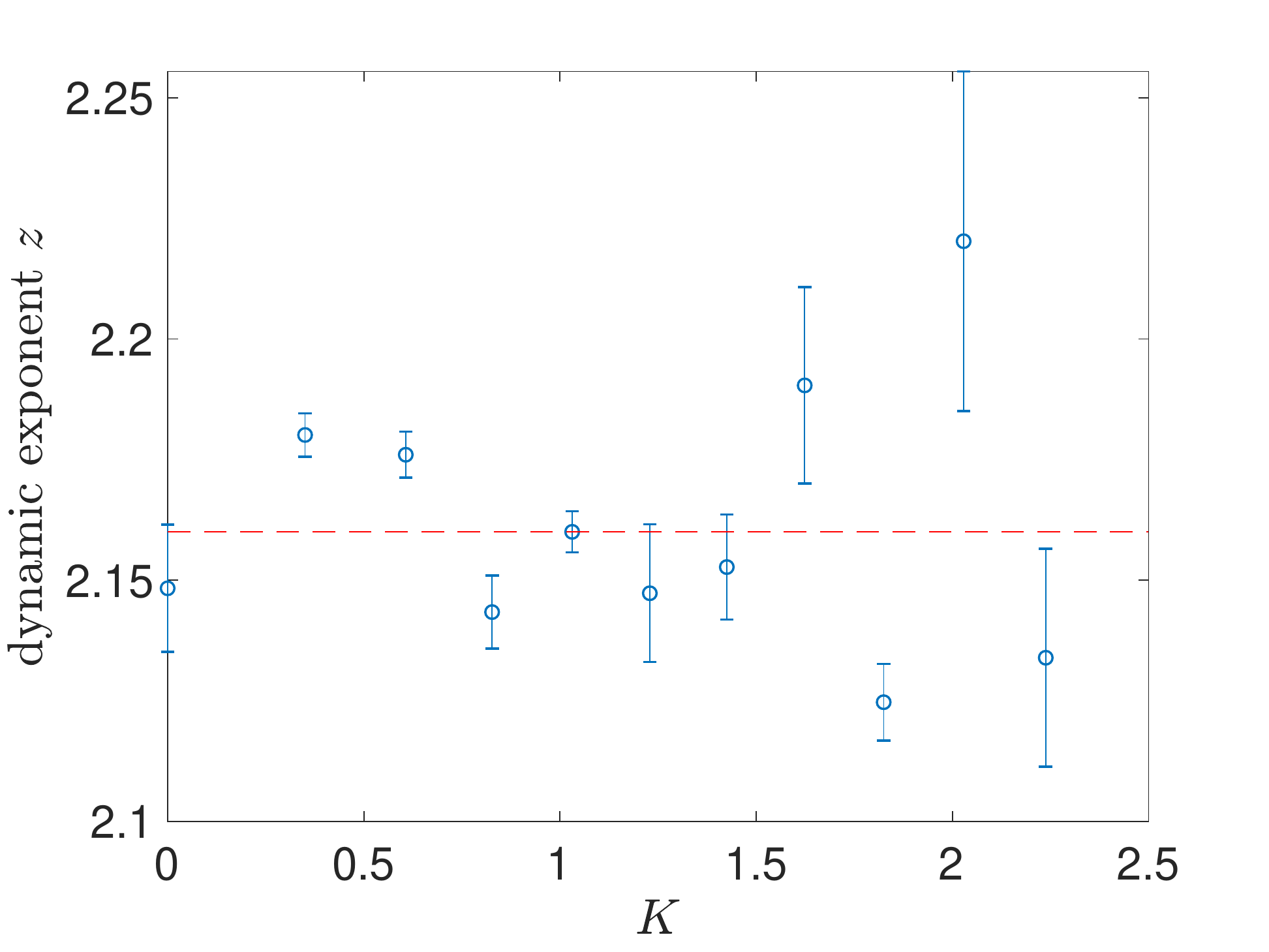}
\caption{\label{fig:z} Dynamic exponent $z$ for the autocorrelation time (see Eq.\ \eqref{eqntau0}) is obtained as function of self-interaction $K$. The dashed horizontal line denotes the non-conserved order parameter Ising universal dynamic exponent, $z=2.16$. 
}
\end{figure}

Figure \ref{fig:z} shows $z(K)$ as a function of $K$ obtained from fits of the form $\tau = A L^{z}$.   The values of $z(K)$ are all reasonably close to the accepted dynamic exponent of the 2D Ising model with non-conserved order parameter, $z=2.16$ \cite{dy_ex3,dy_ex1,dy_ex2}, with no discernable trends in $K$. We believe that the  kinetic Ising model with self-interaction is in the dynamic Ising universality class with a non-conserved order parameter (Model A of \cite{Halperin}) and that the deviations in the measurements from $z(K)=z(0)$ result from a combination of finite-size corrections and statistical errors. This conclusion is in agreement with the result that  systems with local dynamics and a non-conserved  order parameter are in universality class of Ising models \citep{Jayaprakash,Schmittmann}.

\begin{figure}
\centering
\includegraphics[width = .5\textwidth]{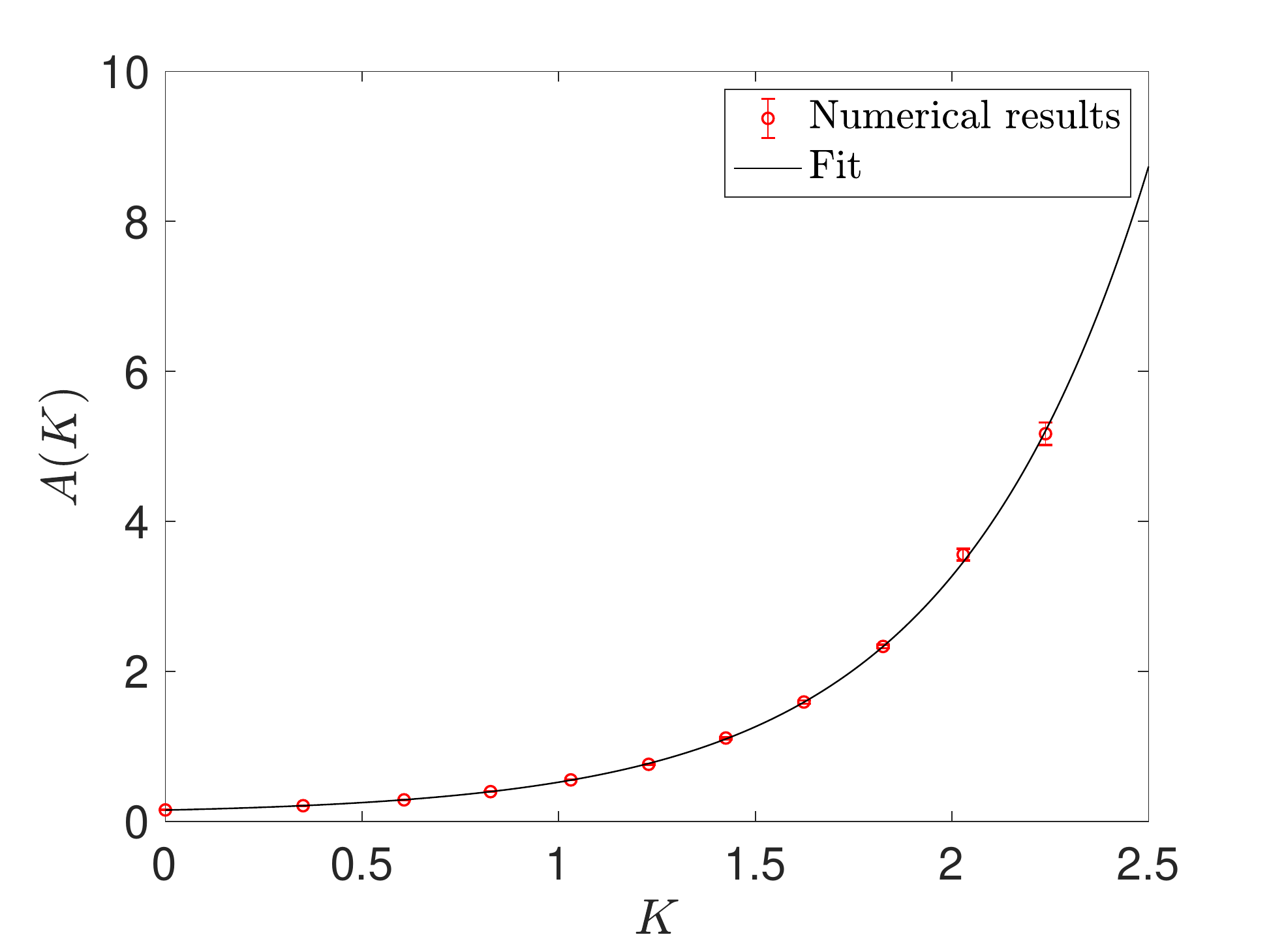}
\caption{\label{fig:amplitude}The amplitude $A(K)$ of the autocorrelation time divergence at criticality (see Eq.\ \eqref{eqntau1}) is plotted as function of $\mem$ for sequential updating. 
%Numerical results of $A(K)$ with error bars are obtained by calculating autocorrelation times \eqref{eqntau1} for lattice sizes $L=10-100$ at critical temperature. 
The solid line is the fit $A(K) = 0.09[1+0.64 \exp(2 K)]$. 
%is  with the numerical data (chi-square per d.o.f \ $0.99$).
}
\end{figure}

We now assume $z$ takes the universal value, $2.16$ and investigate the prefactor, $A(K)$. The autocorrelation time $\tau$ is fitted to the form, 
\begin{equation}\label{eqntau1}
\tau(K,L) = A(K) L^{2.16} + B(K) ,
\end{equation}
where the constant $B(K)$ is included to improve the quality of the fit. %\textcolor{red}{can we say that the quality of the fit is good for all K?}
%\textcolor{orange}{($\chi^2/d.o.f$ lies in between 0.5 and 3.2 for all $K$)}
% The amplitude for a given value of $K$ is obtained by fitting $\tau$ values as function of $L$. 
% Fit of the form \eqref{eqntau1} for $K=1.23$ is shown as solid line in Fig.\ \ref{fig:zfit} ($\chi^2 / d.o.f = 1.49$) and it is clear that the expression fits the obtained data very well . 
The numerical results for the amplitude function $A(K)$ are obtained for various values of $K$ and are shown for sequential updating in Fig.\ \ref{fig:amplitude}. The data can be fit to the form,
\begin{equation}\label{eqntau2}
A(K)=a[1+ b \exp(2 K)],
    %A(K) = 0.09\Big(1+0.64 \exp(2 K))\Big),
\end{equation}
with the best fit parameters, $a=0.09$ and $b=0.64$ ($\chi^2/\makebox{d.o.f} = 0.99$).  The fit is shown as a solid line in the figure.  The factor $\exp(2 K)$ is the time scale between spin flips for an isolated spin with self-interaction and this time scale apparently controls the asymptotic behavior of $A(K)$.  Although the above results are for sequential dynamics, in the limit of large $K$, when the time scale for spin flips is much larger than the number of spins, parallel and sequential dynamics are expected to have the same dynamical properties.

\section{Equilibrium  states }
\label{sec:equilibrium}

The kinetic Ising models with self-interaction ultimately reach an equilibrium state, which can be written in the form of a Gibbs distribution with a dimensionless Hamiltonian, $\mathcal{H}$ so that the probability of a configuration $\alpha$ is given by $P(\alpha) = \exp[-\mathcal{H(\alpha)}]/Z$ with $Z$ the partition function.  This probability and its associated Hamiltonian must satisfy global balance with respect to the transition probabilities,
\begin{equation}
    \sum_\gamma \left[P(\gamma) A(\gamma \rightarrow \alpha) - P(\alpha) A(\alpha \rightarrow \gamma) \right] = 0,
\end{equation}
for all $\alpha$.  
For the case of parallel dynamics, the Hamiltonian satisfies detailed balance  and was obtained in Ref.\ \citep{CIRILLO201436}. The derivation and resulting Hamiltonian is presented in Sec.\ \ref{parallel}.  The Hamiltonian for the sequential case is discussed in Sec.\ \ref{sequential}.  It does not satisfy detailed balance and we have only been able to obtain it as an expansion in powers of $J$. For both parallel and sequential cases, the Hamiltonians contain more than nearest-neighbor, two-spin interactions. 

\subsection{Parallel dynamics} \label{parallel}

From Eqs.\ (\ref{eq:heatbath}) and (\ref{eq:paralleldynamics}), the transition probability for parallel dynamics can be written as,
\begin{equation}
A(\alpha \rightarrow \gamma) = \prod_{i} \frac{\exp(J h_i^\alpha S_i^\gamma+\mem S_i^\alpha S_i^\gamma)}{2\cosh(J h_i^\alpha +\mem S_i^\alpha )},
\end{equation}
so that  
to satisfy detailed balance we required that, 
\begin{equation}\label{ratio}
    \frac{P(\gamma)}{P(\alpha)}= \prod_i \frac{\cosh(J h_i^\gamma +K S_i^\gamma)}{\cosh(J h_i^\alpha +KS_i^\alpha)},
\end{equation}
and the Hamiltonian $\mathcal{H}$ can be written as \citep{Cirillo_parallel_detail_balance}, 
\begin{equation}\label{eqn3}
- \mathcal{H} = \sum_i  \log [\cosh (J h_i + K S_i ) ].
\end{equation}
From Eq. \eqref{ratio}, we should note that in the limit of $K \rightarrow \infty$, 
\begin{equation*}
     \frac{P(\gamma)}{P(\alpha)} = \prod_i \frac{\cosh(J h_i^\gamma S_i^\gamma+K)}{\cosh(J h_i^\alpha S_i^\alpha +K)} \rightarrow \prod_i \frac{\exp(J h_i^\gamma S_i^\gamma)}{\exp(J h_i^\alpha S_i^\alpha)},
\end{equation*}
which implies that, 
\begin{equation}\label{eqn4}
- \mathcal{H} \rightarrow J\sum_i  h_i S_i = 2J\sum_{\langle i,j \rangle1}S_i S_j,
\end{equation}
where the notation $\sum_{\langle \rangle 1}$ indicates a sum over all nearest-neighbor pairs.
Thus, in this limit the Hamiltonian reduces to the standard Ising model with interaction strength twice that of the original kinetic model. This fact has been proved in Ref.\ \citep{Da2012}.
In case of $K \rightarrow -\infty$ we have that, 
\begin{equation}
- \mathcal{H} \rightarrow -J\sum_i h_i S_i = -2J\sum_{\langle i,j \rangle1 }S_i S_j,
\end{equation}
showing that the system behaves as an antiferromagnetic Ising model, again with twice the strength of the interaction in original dynamical model.

The expression for the Hamiltonian, Eq. \eqref{eqn3}, can be expanded as a sum of all products of each spin and its four nearest neighbors with various couplings, \begin{equation}\label{eqn5}
    - \mathcal{H} =  J_1 \mathlarger{\sum}_{\langle i,j \rangle 1} S_i S_j+J_2 \mathlarger{\sum}_{\langle i,j \rangle 2} S_i S_j+J_3 \mathlarger{\sum}_{\langle i,j \rangle 3} S_i S_j +T_4  \mathlarger{\sum}_{\langle i,j,k,l \rangle \perp} S_i S_j S_k S_l + F_4 \mathlarger{\sum}_{\langle i,j,k,l \rangle \diamond}S_i S_j S_k S_l + C
\end{equation}

\begin{figure}
\centering
\includegraphics[width = .5\textwidth]{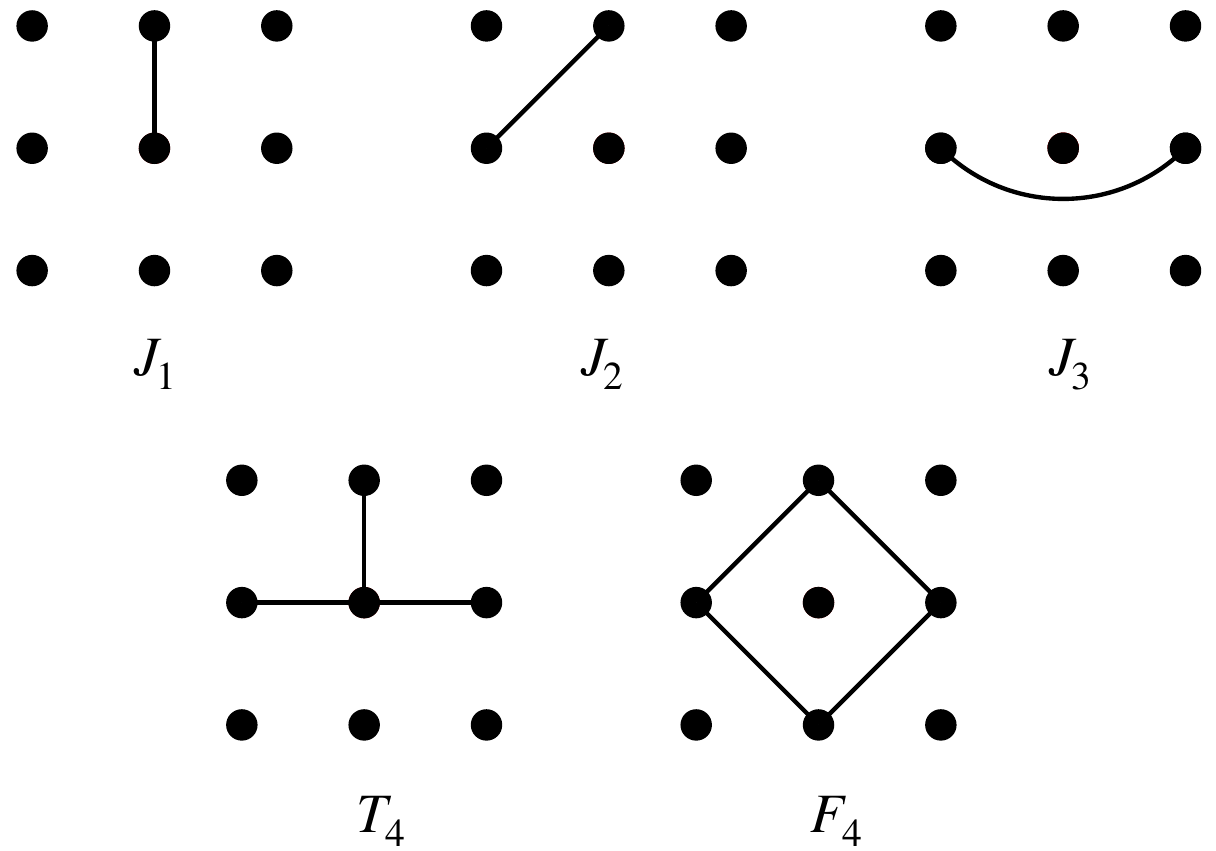}
\caption{\label{fig:lattice} The two-spin couplings, $J_1$, $J_2$ and $J_3$ couple nearest-neighbor, second-nearest-neighbor and third-nearest-neighbor spins, respectively. The four-spin coupling, $T_4$ couples a spin with three of its four neighboring spins. The four-spin coupling, $F_4$ couples the four neighbors of a central spin. All interactions include all rotations and translations of the pictured couplings. }
\end{figure}

 The definitions of the three two-spin couplings and two four-spin coupling allowed by symmetry are illustrated in the Fig.\ \ref{fig:lattice}. 
 The sum $\sum_{\langle \rangle 1}$ is over all pairs of nearest-neighbor spins, whereas $\sum_{\langle \rangle 2}$ and $\sum_{\langle \rangle 3}$ are sums over second- and third- nearest-neighbor spins with coupling constants $J_2$ and $J_3$, respectively.
 $\sum_{\langle \rangle \perp}$ and $\sum_{\langle \rangle \diamond}$ are sums over four spins as shown in Fig.\ \ref{fig:lattice}.  The expressions for the dimensionless couplings $\{J_1,J_2,J_3,T_4,F_4\}$ can be obtained as function of the dynamical interactions $J$ and $K$ by comparing Eqs.\ \eqref{eqn3} and\ \eqref{eqn5} and are given below \citep{CIRILLO201436}:
 \begin{align}
J_1 &=  \frac{1}{4}\log \frac{\cosh(2J + K)}{\cosh(2J - K)} + \frac{1}{8} \log \frac{\cosh(4J + K)}{\cosh(4J - K)} \label{parJ1},\\
T_4 &=   \frac{1}{16} \log \frac{\cosh(4J + K)}{\cosh(4J - K)} -\frac{1}{8}\log \frac{\cosh(2J + K)}{\cosh(2J - K)}\label{parT4}, \\
J_2 &= \frac{1}{8} \log [\cosh(4 J + K)  \cosh(4 J - K)]-\frac{1}{4}\log \cosh(K) \label{parJ2}, \\
J_3 &= \frac{1}{16} \log [\cosh(4 J + K)  \cosh(4 J - K)]-\frac{1}{8}\log \cosh(K)\label{parJ3} , \\
F_4 &= \frac{3}{8} \log \cosh(K)-\frac{1}{4}\log [\cosh(2 J + K)  \cosh(2 J - K)]+\frac{1}{16} \log [\cosh(4 J + K)  \cosh(4 J - K)]\label{parF4} .
\end{align}
 
\begin{figure}
\centering
\includegraphics[width = .5\textwidth]{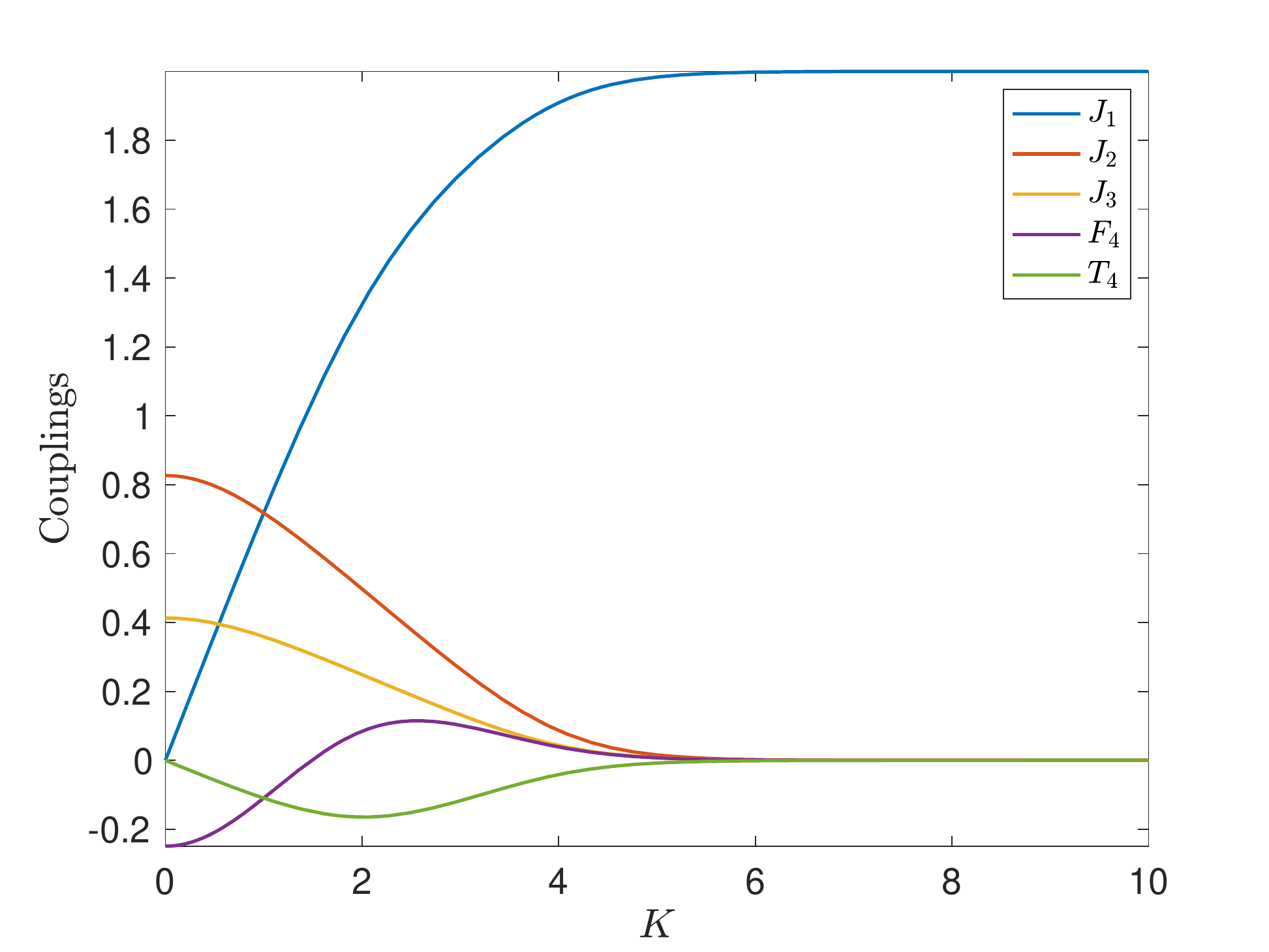}
\caption{\label{fig:parallel}The couplings for parallel dynamics are plotted for $J = 1$ as a function of self-interaction, $\mem$ as given in Eqs.\ \eqref{parJ1}--\eqref{parF4}. For large $\mem$, $J_1 \rightarrow 2J$ while the other couplings vanish. Note that $J_1$ and $T_4$ are odd functions of $\mem$ while the other couplings are even.}
\end{figure}

 The couplings are plotted for $J=1$ as a function of $K$ in Fig.\ \ref{fig:parallel}.  The couplings $\{J_2,J_3,F_4\}$ given in Eqs.\ \eqref{parJ2}-\eqref{parF4} are even in $K$ whereas the couplings $\{J_1,T_4\}$ given in Eqs.\ \eqref{parJ1} and \eqref{parT4}  are odd revealing  antiferromagnetic behavior for negative values of $K$. 

The couplings are non-trivial even when the self-interaction is zero and are discussed in the Sec.\ \ref{zero case}. As $K$ increases, the first nearest neighbor coupling $J_1$ increases to twice the value of $J$ whereas the other couplings decrease to zero.

\subsubsection{Weak self-interaction regime} \label{zero case}

When the self-interaction is zero ($K = 0$), the couplings $J_1$ and $T_4$ given in Eqs.\ \eqref{parJ1} and \eqref{parT4} are zero. The non-zero couplings $J_2$, $J_3$ and $F_4$ given in Eqs.\ \eqref{parJ2}-\eqref{parF4} produce interactions only within a single sublattice. Hence, for  $\mem=0$ the entire system can be viewed as two independent systems, each on one of the  sublattices and with two-spin and four-spin couplings \citep{CIRILLO201436,louis_thesis}. 

It is interesting to note that the values of these two- and four-spin couplings given in Eqs.\ \eqref{parJ2}-\eqref{parF4} at $\mem=0$ are identical to the expressions for the couplings generated when the real space renormalization group decimation scheme \citep{Wilson_kondo, kadanoff_RG} is applied once to the nearest-neighbor Ising model with nearest-neighbor interaction strength $J$. This observation explains why the parallel update Ising model with $\mem=0$ has a critical point at $J_c^{(p)}(0)=J_c(0)$, the critical point of the standard Ising model. A rigorous proof of this fact is given in \citep{louis_thesis}. 

The behavior of the parallel update model for $\mem=0$ differs from the standard Ising model because of the presence of two ``checkerboard'' ground states and, at finite but low temperature, both ferromagnetic and checkerboard ordered phases. The two checkerboard phases each display period-two oscillations between the two antiferromagnetically ordered states and differ in the phase of this oscillation.  Furthermore, the four ordered phases each have fluctuations of all four kinds.  For example, the ferromagnetic phases will contain checkerboard domains.  

\begin{figure}
\centering
\includegraphics[width = .3\textwidth]{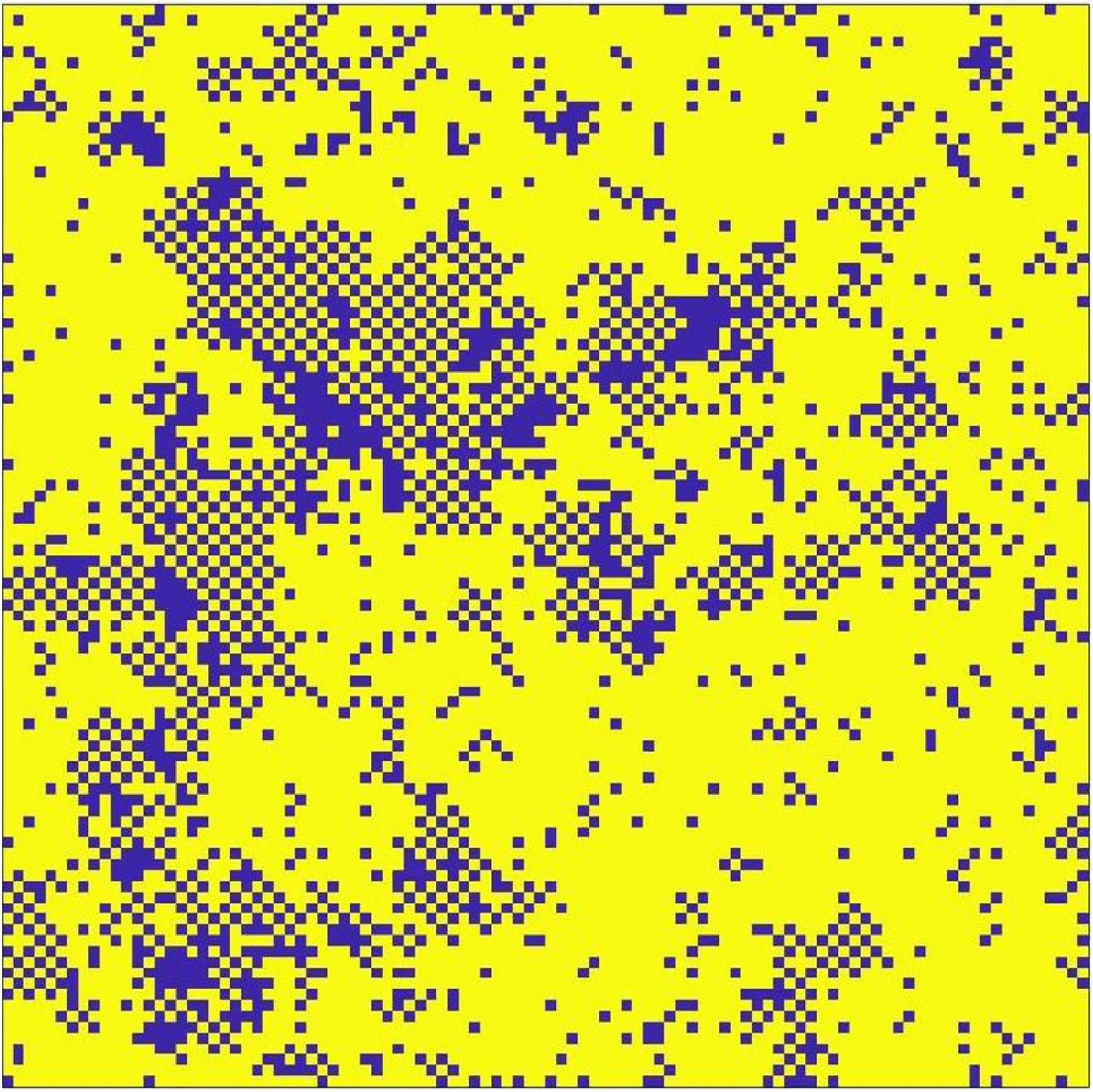}
\caption{\label{fig:checkerboard} Snapshot of a $100\times100$ lattice in equilibrium with parallel updating for $K=4\times 10^{-4}$ at the critical coupling $J_c^{(p)}(K) = 0.4360$. The presence of large antiferromagnetic clusters result from weak nearest-neighbor interactions (set by $J_1$) and strong next-nearest-neighbor interactions (set by $J_2$). }
\end{figure}

As the value of $K$ increases, the couplings  that are responsible for interactions between sublattices, $J_1$ and $T_4$, increase. Hence, the sublattices are positively correlated and only the two ferromagnetic ground states survive \citep{CIRILLO201436,Cirillo_parallel_detail_balance}.  However, at finite temperature and $\mem$ small, checkerboard domains remain common. Figure \ref{fig:checkerboard} shows a snapshot of the equilibrium state for $\mem=4\times10^{-4}$ at the critical coupling, which displays large checkerboard domains.  Similarly, for $\mem$ small and negative, and $J$ near its critical value one finds  large ferromagnetic domains.

As can be seen in Fig.\ \ref{fig:phase_plot} and the associated fit, Eq.\ \eqref{par_fit}, the critical curve, $J^{(p)}_c(K)$ appears to approach the $\mem=0$ critical point, $(J_c(0),0)$ with infinite slope.  The phase diagram is even in $\mem$ so there is a corresponding critical curve approaching $(J_c(0),0)$ from negative $\mem$. We now give a heuristic argument based on finite-size scaling considerations for the singular behavior of these critical curves near $(J_c(0),0)$.     Near the critical point $(J_c(0),0)$ there are large fluctuations in both the ordinary (ferromagnetic) magnetization and the  staggered (antiferromagnetic) magnetization. These fluctuations are quantified, respectively, by the ordinary susceptibility, $\chi$, and the staggered susceptibility, $\chi_s$, both of which diverge at $(J_c(0),0)$.  Along the critical curve, $J_c^{(p)}(\mem)$ for $\mem>0$, the ordinary susceptibility diverges while along the corresponding critical curve for negative $\mem$ the staggered susceptibility diverges.        

Consider a geometric mean susceptibility, $\chim$ defined as $\chim = \sqrt{\chi \chi_s}$.   This mean susceptibility for systems of size $L$ with periodic boundary conditions and with $L$ even is expected to diverge as $L^{\gamma/\nu}$ at $(J_c(0),0)$.  For $\mem \neq 0$ along the two critical curves, $\chim$ is expected to have a weaker finite-size divergence, $L^{\gamma/2\nu}$.  We now hypothesize that the finite-size scaling behavior of $\chim$ as a function of $L$, $J$ and $\mem$ takes the form, 
\begin{equation}
\chim (J, \mem) \approx L^{\gamma/\nu} f((J-J_c(0))L^{a}, \mem L^b).
\end{equation}
The qualitative behavior of the finite-size scaling function $f(x,y)$ is that it has a maximum along the line $y=\mem=0$ near $x=0$.  Furthermore,  $f$ will decrease in all directions away from the maximum but with two ridge lines near the two critical curves.    These ridge lines are expected to take the form $x \sim  -(\pm y)^c$ near maximum for the $\pm \mem >0$ critical curves, respectively.  Given this picture, we must have,
\begin{equation}
    (J_c^{(p)}(K)-J_c(0))L^{a} \sim  - (|\mem| L^b))^c,
\end{equation}
so that $c=a/b$.

What are the scaling exponents $a$ and $b$?  Since near $J_c(0)$, $J$ controls the strength of the relevant interactions linearly along the $\mem=0$ line, it is reasonable to assume the usual finite-size scaling exponent, $a=1/\nu$, so, in the present case, $a=1$. 

The exponent $b$ determines the rate of decrease of $f$ for a fixed $K$ as $L$ increases.  We hypothesize that this decrease is dominated by the suppression of the ``wrong'' kinds of fluctuations.  For $\mem >0$ these are staggered magnetization fluctuations and for $\mem <0$ ordinary magnetization fluctuations.  For small $\mem$ the coupling $J_1$ between sublattices increases linearly in $\mem$,  specifically (see Eq.\ \eqref{parJ1}), $J_1=2 J \mem + O(\mem^3)$.  Thus the energetic cost of the wrong kind of fluctuation of length scale $\ell$ is given by $2 J \mem \ell^2$ and the wrong kind of fluctuation is  suppressed  when this energy exceeds $k_B T_c$.  The maximum size of these fluctuations is, therefore, $\sqrt{k_B T_c/2 \mem}$ and the associated finite-size scaling exponent for $\mem$ is  $b=2$.  This estimate ignores the presence of ferromagnetic fluctuations embedded in larger checkerboard regions so that $b$ is expected to be somewhat less than two.

The conclusion of these arguments is that the critical lines approach the $\mem=0$ critical point with a singularity,
\begin{equation}
 J_c^{(p)}(K) -J_c(0) \sim -|K|^{c},  
\end{equation}
and that $c \gtrapprox 1/2$.

%%%%%%binder cumulant
We obtain numerical results for the critical line for small $\mem$ using the Binder cumulant method.  Care must be taken in using  Binder cumulants for small $\mem$ because the sublattices are nearly uncoupled. Indeed, for $\mem=0$ we now show that the Binder cumulant is half its usual value. Let $M_1, M_2$ be the magnetization of the even and odd sublattices, respectively. The sublattice binder cumulant is defined by,
\begin{equation}
\label{eq:subbinder}
    \U = 1-\frac{ \langle M_1^4 \rangle }{3 \langle M_1^2 \rangle ^2}.
\end{equation}
At the  $(J_c(0),0)$ critical point each sublattice behaves as an independent critical Ising model so $\U$ will take the  Ising critical value for a square lattice with periodic boundary conditions, $U^* \approx 0.61$ \citep{Selke2006}.
The full lattice magnetization, $M$ is given by the sum of the two sublattice magnetizations, $M=M_1+M_2$. 
Expanding the moments of the full magnetization in terms of the sublattice magnetizations yields,
\begin{gather}
\langle M^2 \rangle = \langle M_1^2\rangle+ \langle M_2^2\rangle+ \langle 2 M_1M_2\rangle \\
\langle M^4 \rangle = \langle M_1^4\rangle + \langle M_2^4\rangle + \langle 4M_1^3M_2\rangle + \langle 4M_2^3M_1\rangle + \langle 6M_1^2M_2^2\rangle.
\end{gather}
Since $M_1, M_2$ are identically distributed,   independent random variables we have, 
\begin{gather} 
\langle M^2 \rangle = 2 \langle M_1^2 \rangle \\
\langle M^4 \rangle = 2 \langle M_1^4 \rangle +6 \langle M_1^2 \rangle ^2. 
\end{gather}
Hence, using the definition of the full lattice Binder cumulant, Eq.\ \eqref{binder}, we have,
\begin{equation}\label{binder_half}
U = \frac{1}{2}-\frac{ \langle M_1^4 \rangle }{6 \langle M_1^2 \rangle ^2} = \frac{1}{2} \U. 
\end{equation}

We see that the Binder cumulant of the full lattice for the parallel dynamics, $U$ is half the sublattice Binder cumulant $U_s$ and thus, at criticality it is $(1/2)U^*$.  On the other hand, for any $\mem>0$ and for sufficiently large $L$ we expect the full lattice Binder cumulant to take the usual Ising value, $U^*$.  Thus, large finite-size corrections are expected for both the full lattice Binder cumulant and also the sublattice Binder cumulants when $\mem$ is small. The reason for these corrections is the weak coupling between sublattices and the associated presence of large checkerboard domains. We previously estimated that the maximum domain size  scales as $\sqrt{J_c(0)/2 \mem}$.  For example, if $K=4 \times 10^{-4}$ this size is about $20$, which is comparable to our system sizes. 

\begin{figure}
\centering
\includegraphics[width = .5\textwidth]{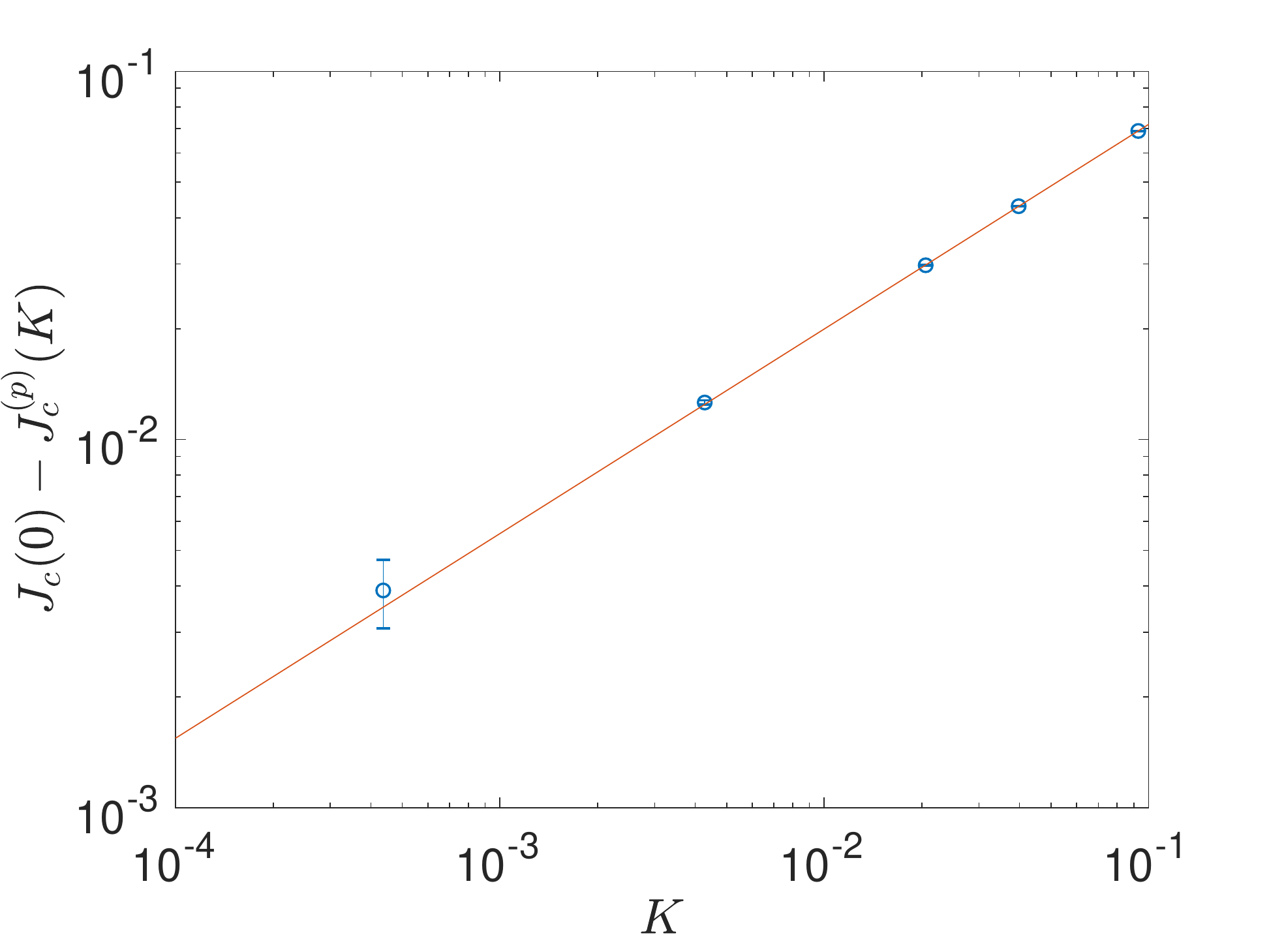}
\caption{\label{fig:loglog}Numerical results for the deviation of the critical dynamic coupling from the Ising critical value, $J_c(0) - J_c^{(p)}(K)$ as a function of $\mem$ for parallel updating. The solid line is a fit to a power law, by Eq.\ \eqref{singular_eqn}, with exponent $c=0.555$.}
\end{figure}

We have investigated the singular behavior of the critical curve through numerical simulation of five values of $\mem$ less than  $0.1$.  Here we determine the critical couplings from both the sublattice Binder cumulant and the full lattice Binder cumulant.  We use a pair of large system sizes, $L=100$ and $120$, ($L=30$ and $60$ are used elsewhere) to obtain the crossing of the Binder cumulant curves.  The reported values of $J_c^{(p)}(K)$ are obtained by averaging the full lattice and sublattice results in the hope that this procedure will reduce finite-size corrections.   Figure \ref{fig:loglog} shows the numerical results for the critical dynamic coupling $(J_c(0) - J_c^{(p)}(\mem))$ as a function of $\mem$.  The error bars shown in the figure are either the difference of the sublattice and full lattice values or the maximum of the statistical errors, whichever is larger. The larger error bars for the three smallest $\mem$ data points are the result of the large difference between results from the two Binder cumulants and reflect an estimate of the size of systematic errors.

A fit of the data to the form,
\begin{equation}\label{singular_eqn}
    J_c(0)-J_c^{(p)}(K) = d K^{c} ,
\end{equation}
yields $c=0.555\pm0.001$  and $d=0.26$ and  ($\chi ^2 /$ d.o.f = 0.83).  This fit is shown as the solid line in Fig.\ \ref{fig:loglog}.   The numerical result for $c$ are consistent with the theoretical arguments presented above.  We note that the small error bar on $c$ may be misleading because it is set by the data points with small error bars but relatively large values of $\mem$, which may hide corrections to scaling in the simple fit, Eq.\ \eqref{singular_eqn}.  Independent fits of the sublattice and full lattice results using their respective statistical errors also yield $c \approx 0.55$ but the quality of these fits is poor.  It would be interesting to carry out a thorough study of the exponent $c$ and, more generally, the finite-size scaling behavior near the $\mem=0$ critical point. 

%For values of $\mem$ less than $2.2 \times 10^{-3}$ (\textcolor{red}{this value of K corresponds to lattice size $L=10$}) we therefore determine the critical coupling from the the sublattice Binder cumulant. \textcolor{red}{(I actually used Binder cumulant of entire lattice with higher lattice sizes.)}
% For two-dimensional Ising model, correlation length $\zeta$ 
% \begin{equation}\label{scaling2}
%     \zeta \sim (T - T_c)^{-\nu}
% \end{equation}
% where the exponent $\nu =1$. Checkerboard being one of the ground states, $\zeta_{ch}$ follows the Eq.\ \eqref{scaling2}.   Combining Eqs.\ (\ref{scaling1}, \ref{scaling2}), we get the relation between $K$ and $T-T_c$ as %\textcolor{red}{need more justification for combining these equations}
% \begin{equation}
%     T-T_c \sim K^{1/2}.
% \end{equation}
%Figure \ref{fig:loglog} shows the numerical results of relative nearest neighbor coupling $J_c(0) - J_c^{(p)}(K)$ for small values of self-interaction $K$.  For $K=4 \times 10^{-4}$ we obtain $J_c^{(p)}(K)$ with system sizes $L=100,120$ to reduce finite size effects while $L=30, 60$ for higher values of $K$. 

\subsection{Sequential dynamics} \label{sequential}
% The new state $\gamma$ differs at most by one spin from the old state $\alpha$ in sequential dynamics and let the spin that is different be $S_i$. 
From Eqs.\ \eqref{eq:heatbath} and \eqref{acceptanceratio}, the transition probability when $\alpha$ and $\gamma$ differ by a single spin flip at site $i$ ($\chi_1(\alpha,\gamma,i)=1$) is given by,
\begin{equation}\label{transitionprobabilityseq}
A(\alpha \rightarrow \gamma) =  \frac{\exp(J h_i^\alpha S_i^\gamma+\mem S_i^\alpha S_i^\gamma)}{2 N\cosh(J h_i^\alpha +\mem S_i^\alpha)}.
\end{equation}

% The above equation holds if $\chi_1(\alpha,\gamma,i)=1$ and is not defined otherwise. Neighborhood of $S_i$ remain same for the state $\alpha$ and $\gamma$, i.e.,  $h_i^\alpha = h_i^\gamma$.

First consider the three limiting cases, $K=0$ and $K \rightarrow \pm \infty$ (holding $J$ fixed), for which detailed balance holds.  Note that for sequential dynamics, the neighborhood of $S_i$ is the same for states $\alpha$ and $\gamma$ so  $h_i^\alpha = h_i^\gamma \equiv h_i$.  The detailed balance condition is,
\begin{equation}\label{detailedbalanceseq}
%  \frac{P_\gamma}{P_\alpha} =  \frac{\exp(J h_i^\alpha S_i^\gamma)} { \cosh( J h_i^\alpha  + K S_i^{\alpha})}\frac{ \cosh(J h_i^\gamma  +  K S_i^{\gamma})}{\exp( J h_i^\gamma S_i^\alpha)}.
 \frac{P(\gamma)}{P(\alpha)} =  \frac{\exp(J h_i S_i^\gamma)} { \cosh( J h_i  + K S_i^{\alpha})}\frac{ \cosh(J h_i  +  K S_i^{\gamma})}{\exp( J h_i S_i^\alpha)}.
\end{equation}
For $K=0$ we have standard heat-bath dynamics, 
\begin{equation}
\frac{P(\gamma)}{P(\alpha)} = \exp[ J h_i (S_i^\gamma-S_i^\alpha)],
\end{equation}
and the equilibrium distribution is that of the standard nearest-neighbor Ising model, $P(\alpha) \propto \exp[(J/2) \sum_i h_i S_i]$. In the limit $K \rightarrow +\infty$,  
% \begin{equation*}
%   \cosh({J h_i + KS_i^\alpha}) 
%  \rightarrow  \frac{\exp(J h_i S_i^\alpha + K) }{2}
% \end{equation*}
% and 
the ratio of probabilities becomes,
\begin{equation}\label{highK}
 \frac{P(\gamma)}{P(\alpha)}  \rightarrow  \exp[2 J h_i (S_i^\gamma-S_i^\alpha)]
\end{equation}
showing that the equilibrium distribution is that of the standard Ising model with twice the nearest-neighbor coupling ($J \rightarrow 2J$).  Note, however, that when $K \rightarrow + \infty$ the time scale for a spin flips diverges so that if the limit is taken with the number of sweeps (and system size) held fixed then the spin configuration is fixed.  Finally, when $K \rightarrow -\infty$, we have that
% \begin{equation*}
%   \cosh({J h_i + KS_i^\alpha}) 
%  \rightarrow  \frac{\exp(-J h_i S_i^\alpha - K) }{2}
% \end{equation*}
$P(\gamma)/P(\alpha)  \rightarrow  1$
% the ratio of probabilities is,
% \begin{equation}\label{lowK}
%  \frac{P_\gamma}{P_\alpha}  \rightarrow  1
% \end{equation}
showing that the model behaves as non-interacting (infinite temperature) system.

For $0<\mem<\infty$  the equilibrium distribution does not satisfy detailed balance with respect to the dynamics and one must solve the global balance equations,
% The transition probability for Ising model with self-interaction is given by Eq.\ \eqref{transitionprobabilityseq}, 
% \begin{equation}
%     A(\alpha \rightarrow \gamma) = \frac{\exp(J h_i^{\alpha} S_i^{\gamma} + K S_i^{\alpha}S_i^{\gamma})}{2 \cosh (J h_i^{\alpha} + K S_i^{\alpha})}.
% \end{equation}
% For global balance, the equilibrium  probability $P(\gamma)$ that the system will be in state $\gamma$ satisfies the equation,
\begin{equation}\nonumber
    \sum_\gamma\left[ P(\gamma) A(\gamma \rightarrow \alpha) - P(\alpha) A(\alpha \rightarrow \gamma)\right] = 0,
\end{equation}
which can be rewritten as,
\begin{equation}
    \sum_\gamma \left[ \frac{P(\gamma)}{P(\alpha)} A(\gamma \rightarrow \alpha) -  A(\alpha \rightarrow \gamma) \right] = 0.
\end{equation}

Let $\alpha$ be an arbitrary initial, $\alpha = \{ S_j^\alpha\}$ and let $\alpha(i)\equiv \gamma$ refers to the state obtained from $\alpha$ by flipping spin $S_i$.  From Eq.\ \eqref{transitionprobabilityseq},  global balance  takes the form,
% Since for sequential dynamics, the next state $\gamma$ differs at most by one spin, so global balance takes the form,
\begin{equation}\nonumber
 \sum_i \left[ \frac{P(\alpha(i))}{P(\alpha)} \frac{\exp(J h_i^\alpha S_i^\alpha - K)}{2 \cosh(J h_i^\alpha - K S_i^\alpha)} - \frac{\exp(-J h_i^\alpha S_i^\alpha - K)}{2 \cosh(J h_i^\alpha + K S_i^\alpha)}\right] = 0,
\end{equation}
which can be simplified to,
\begin{equation}\label{eq1}
 \sum_i \left[\frac{P(\alpha(i))}{P(\alpha)} \frac{\exp(J h_i S_i)}{\cosh(J h_i - K S_i)} - \frac{\exp(-J h_i S_i)}{\cosh(J h_i + K S_i)}\right] = 0.
\end{equation}
Superscripts are dropped in the above equation and what follows since all spins belong to configuration $\alpha$. 

\begin{figure}
    \centering
    \includegraphics[width = 0.7\textwidth]{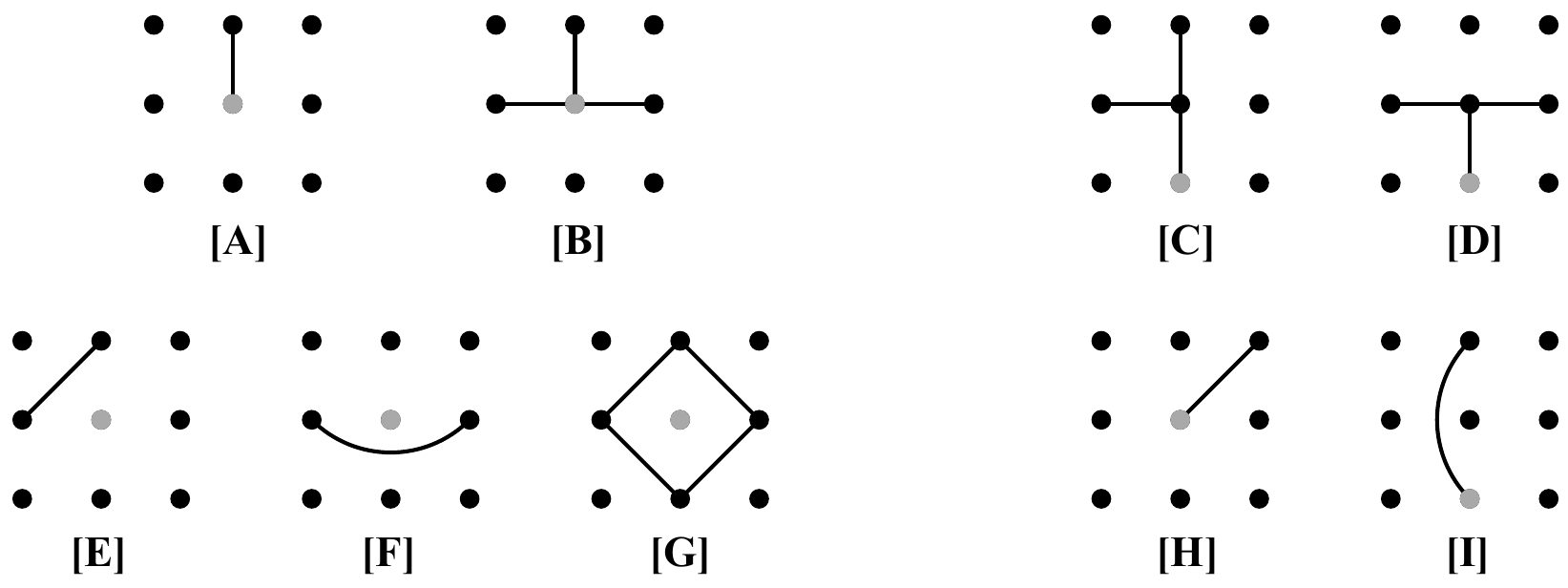}
    \caption{The dynamical interactions in the expansion of the transition probabilities are shown on the left while additional interaction appearing in the Hamiltonian are shown on the right. All interaction types are labelled with boldface letters. The grey lattice site represents the flipped spin $i$. All the interactions include all distinct rotations around the grey site and, in the case of [$\textbf{C}$], reflection around the vertical axis. There are 4 equivalent interactions for type [$\textbf{A}$], 4 for [$\textbf{B}$], 8 for [$\textbf{C}$], 4 for [$\textbf{D}$], 4 for [$\textbf{E}$], 2 for [$\textbf{F}$], 1 for [$\textbf{G}$], 4 for [$\textbf{H}$] and 4 for [$\textbf{I}$].  }
    \label{couplings}
\end{figure}

The transition probability factors in Eq.\ \eqref{eq1} can be expanded in terms of spin couplings involving the flipped spin, $S_i$,  and its four nearest neighbors, %\textcolor{red}{use a, b etc to indicate local couplings relative to $i$}
\begin{align}
\label{eq:seqexpand0}
   \frac{ \exp ( \pm J h_i S_i)}{\cosh(J h_i \mp K S_i)} = \exp \Big(&\pm J_1 S_i \sum_{\langle j;i\rangle A} S_j  -J_2 \sum_{\langle j,k;i \rangle E} S_j S_k-J_3 \sum_{\langle j,k;i \rangle F} S_j S_k \pm T_4 S_i \sum_{\langle j,k,l;i\rangle B} S_j S_k S_l \nonumber \\ &- F_4 \sum_{\langle j,k,l,m;i\rangle G} S_j S_k S_l S_m + C \Big)
\end{align}
where $\sum_{\langle ...;i\rangle X}$ is summation among the neighbors of spin $i$ of type $X$ as shown in Fig.\ \ref{couplings}. To simplify the notation,  spin couplings are denoted with bracketed, boldface letters as shown in Fig.\ \ref{couplings} and written explicitly here: %\textcolor{red}{should we put square brackets around the boldface letters?}
\begin{align}
    \nonumber [\textbf{A}] &= S_i \sum_{\langle j;i \rangle A} S_j \qquad \quad
    [\textbf{B}] = S_i \sum_{\langle j,k,l;i \rangle B} S_j S_k S_l \quad \quad
    [\textbf{C}] = S_i \sum_{\langle j,k,l;i \rangle C} S_j S_k S_l \quad \quad
    [\textbf{D}] = S_i\sum_{\langle j,k,l;i \rangle D} S_j S_k S_l \\ \nonumber
    [\textbf{E}] &=  \sum_{\langle j,k;i\rangle E} S_j S_k    \qquad \quad
    [\textbf{F}] =  \sum_{\langle j,k;i \rangle F} S_j S_k  \qquad \qquad \quad
    [\textbf{G}] =  \sum_{\langle j,k,l,m;i\rangle G} S_j S_k S_l S_m  \quad \quad
    [\textbf{H}] = S_i \sum_{\langle j;i \rangle H} S_j \qquad \qquad \\ 
    [\textbf{I}] &= S_i \sum_{\langle j;i \rangle I} S_j .
\end{align}
Note that these couplings are implicitly functions of $i$.

In terms of this compact notation, 
\begin{equation}
\label{eq:seqexpand}
   \frac{ \exp ( \pm J h_i S_i)}{\cosh(J h_i \mp K S_i)} = \exp \Big(\pm J_1 [\textbf{A}]  -J_2 [\textbf{E}]-J_3 [\textbf{F}]  \pm T_4 [\textbf{B}] - F_4 [\textbf{G}] + C \Big),
\end{equation}
where $C$ is a constant.

Expanding both sides of the above equation and comparing like terms yields expressions for the coupling coefficients in terms of $J$ and $\mem$,
\begin{align}\label{eq2}
J_1 &=  J + \frac{1}{8}\log \frac{\cosh(2J + K)}{\cosh(2J - K)} + \frac{1}{16} \log \frac{\cosh(4J + K)}{\cosh(4J - K)}, \\
T_4 &=   \frac{1}{16} \log \frac{\cosh(4J + K)}{\cosh(4J - K)} -\frac{1}{8}\log \frac{\cosh(2J + K)}{\cosh(2J - K)}, \\
J_2 &= \frac{1}{8} \log [\cosh(4 J + K)  \cosh(4 J - K)]-\frac{1}{4}\log \cosh(K), \\
J_3 &= \frac{1}{16} \log [\cosh(4 J + K)  \cosh(4 J - K)]-\frac{1}{8}\log \cosh(K), \\
F_4 &= \frac{3}{8} \log \cosh(K)-\frac{1}{4}\log [\cosh(2 J + K)  \cosh(2 J - K)]+\frac{1}{16} \log [\cosh(4 J + K)  \cosh(4 J - K)].
\end{align}
Except for $J_1$ these couplings are identical to the couplings appearing in the Hamiltonian for the equilibrium state of parallel dynamics, Eqs.\ \eqref{parJ1}-\eqref{parF4}.  However, it is important to note that these dynamical coupling coefficients for sequential dynamics are not the same as the coupling coefficients appearing in the sequential Hamiltonian, which we now derive perturbatively in $J$.  We begin by expanding the dynamical coupling coefficients to fourth order in $J$,
\begin{align}
J_1 &= \big[\tanh (K)+1\big] J +\frac{10}{3}  \big[\tanh ^3(K)-\tanh (K)\big] J^3+ O[J^5], \label{eq:4J1}\\
T_4 &=  \big[2 \tanh ^3(K)-2 \tanh (K)\big] J^3 + O[J^5], \\
J_2 &=  \big[2- 2 \tanh ^2(K)\big] J^2+ \big[-16 \tanh ^4(K)+\frac{64 \tanh ^2(K)}{3}-\frac{16}{3}\big] J^4 +O[J^6], \\
J_3 &= \frac{J_2}{2} = \big[ 1-\tanh ^2(K)\big] J^2+ \big[-8 \tanh ^4(K)+\frac{32 \tanh ^2(K)}{3}-\frac{8}{3}\big] J^4 +O[J^6], \label{eq:4J3}\\
F_4 &=  \big[-6 \tanh ^4(K)+8 \tanh ^2(K)-2\big] J^4 + O[J^6].
\end{align}

Inserting the expansion of the transition probability, Eq.\ \eqref{eq:seqexpand}, in the global balance equation, Eq.\ \eqref{eq1}, yields, 
\begin{flalign}
\label{eq:globalexpand1}
  \nonumber  & \sum_i \Big[\frac{P(\alpha(i))}{P(\alpha)} \exp \big(J_1 [\textbf{A}]  -J_2 [\textbf{E}]-J_3 [\textbf{F}]+ T_4 [\textbf{B}] - F_4 [\textbf{G}] \big) &\\
    &\qquad \qquad \; - \exp \big(-J_1 [\textbf{A}]  -J_2 [\textbf{E}]-J_3 [\textbf{F}]- T_4 [\textbf{B}] - F_4 [\textbf{G}]  \big) \Big] = 0.&
\end{flalign}

We now make the ansatz that $P(\alpha)$ can be written in Gibbsian form with a Hamiltonian potentially containing all even order spin couplings,
\begin{equation}
\label{eq:Hansatz}
      \mathcal{-H} = \tilde{J}_1  \sum_{\langle i,j \rangle 1}S_i S_j + \tilde{J}_2 \sum_{\langle i,j\rangle 2} S_j S_k + \tilde{J}_3 \sum_{\langle i,j\rangle 3} S_j S_k + \tilde{T}_4  \sum_{\langle i,j,k,l\rangle\perp} S_i S_j S_k S_l + \ldots
 \end{equation}
The summations in the terms that are explicitly written  are defined above in Sec.\ \ref{parallel} (see Eq.\ \eqref{eqn5} and Fig.\ \ref{fig:lattice}) and couplings in the sequential Hamiltonian are written with tildes.
% \textcolor{orange}{Even though there can be many higher order interactions in the Hamiltonian, the only possible interactions upto third order in $J$ are the ones that have less than four "bonds" and exists within "cross" spins (for every spin, its four nearest neighbors and itself forms a cross). }
Given this ansatz, the ratio of probabilities is, 
 \begin{equation}\label{eq3}
    \frac{P(\alpha(i))}{P(\alpha)} =\exp\big[-2 \big(\tilde{J}_1 [\textbf{A}] + \tilde{T}_4 [\textbf{B}] + \tilde{T}_4 [\textbf{C}] + \tilde{T}_4 [\textbf{D}] + \tilde{J}_2 [\textbf{H}] + \tilde{J}_3 [\textbf{I}]\big)\big].
\end{equation}
%where $\sum_{\perp b}$, $\sum_{\perp c}$ and $\sum_{\perp d}$ are summation over the interactions of type $\textbf{B}$, $\textbf{C}$ and $\textbf{D}$ as shown in the Fig. \ref{couplings}. 

Inserting this expression 
%for $\frac{P(\alpha(i))}{P(\alpha)}$ 
in the global balance equation, \eqref{eq:globalexpand1}, yields,
\begin{align}
\label{eq:globalbalancebf}
     \nonumber\sum_i &\Big[\exp[(-2 \tilde{J}_1 +J_1) [\textbf{A}] + (-2\tilde{T}_4+T_4) [\textbf{B}] -2\tilde{T}_4 [\textbf{C}]- 2\tilde{T}_4 [\textbf{D}] - 2\tilde{J}_2 [\textbf{H}]- 2\tilde{J}_3 [\textbf{I}]-J_2 [\textbf{E}]-J_3 [\textbf{F}] - F_4 [\textbf{G}]]&\\
    - &\exp(-J_1 [\textbf{A}] - T_4 [\textbf{B}] -J_2 [\textbf{E}]-J_3 [\textbf{F}] - F_4 [\textbf{G}])\Big] = 0&
\end{align}
We now explicitly expand all the couplings as series in the dynamic coupling, $J$. The coefficients of each series are represented by lower case letters and subscripts corresponding to the name of the coupling while the power in $J$ is indicated with a superscript.  Thus, for example, the nearest-neighbor coupling in the dynamical rule is,
\begin{equation}
    J_1(J) =  j_1^1 J  + j_1^2 J^2 + j_1^3 J^3 +\cdots
\end{equation}
where, from Eq.\ \eqref{eq:4J1}, $j_1^1=[\tanh (K)+1]$, $j_1^2 = 0$, and $j_1^3=\frac{10}{3}  [\tanh ^3(K)-\tanh (K)]$.  The corresponding coefficients of the couplings in the Hamiltonian,  $\tilde{j}_1^n$, $\tilde{j}_2^n$, $\tilde{j}_3^n$, $\tilde{t}_4^n$  are the quantities we wish to solve for.

Expanding Eq.\ \eqref{eq:globalbalancebf} to first order in $J$ yields,
\begin{equation}
    \sum_i \Big [(2 \tilde{j}_1^1 - 2 j_1^1)[\textbf{A}]+  2\tilde{t}_4^1 ([\textbf{B}]+[\textbf{C}]+[\textbf{D}])+2\tilde{j}_2^1 [\textbf{H}]+ 2\tilde{j}_3^1 [\textbf{I}]\Big] = 0
\end{equation}
The sums over the four distinct interaction terms, $[\textbf{A}]$, $([\textbf{B}]+[\textbf{C}]+[\textbf{D}])$, $[\textbf{H}]$, and $[\textbf{I}]$ must vanish independently, thus to first order in $J$,
\begin{equation}\label{firstorder}
    \tilde{j}_1^1 = j_1^1=\big[\tanh (K)+1\big],
\end{equation}
while the coefficients $\tilde{j}_2^1$, $\tilde{j}_3^1$ and $\tilde{t}_4^1$ are all zero. Using the solutions of order $J$ and expanding Eq.\ \eqref{eq:globalbalancebf} to second order in $J$ yields,
%\eqref{allorders}
%for order $J^2$, we get
\begin{equation}
    \sum_i\Big[ 2\tilde{j}_1^2 [\textbf{A}]+2\tilde{t}_4^2 ([\textbf{B}]+[\textbf{C}]+[\textbf{D}])+2\tilde{j}_2^2 [\textbf{H}]+ 2\tilde{j}_3^2 [\textbf{I}]\Big] =0,
\end{equation}
so that $\tilde{j}_1^2$, $\tilde{j}_2^2$, $\tilde{j}_3^2$ and $\tilde{t}_4^2$ all vanish. 
% as zero showing that the $J^2$ contributions to $\tilde{J}_1$,$\tilde{J}_2$,$\tilde{J}_3$ and $\tilde{T}_4$ are all zero.

% \textcolor{orange}{At this and higher orders of $J$, products of the interactions appear as part of the exponential expansion and some of those products with their equivalent couplings are shown below. 
% \begin{align}
 %   [\textbf{A}.\textbf{A}] &= 4+2 [\textbf{E}] + 2 [\textbf{F}] \\
 %   [\textbf{A}.\textbf{B}] &= 4[\textbf{G}]+2 [\textbf{E}] + 2 [\textbf{F}] \\
%    [\textbf{A}.\textbf{C}] &= diverges\\
 %   [\textbf{A}.\textbf{A}.\textbf{A}] &= 10[\textbf{A}]+6[\textbf{B}].
%\end{align}
%The above products of interactions do not show up in the solutions as the coefficients of these  products are zero upto third order in $J$ when the solutions of previous orders are applied.}

Note that $2 \sum_i [\textbf{B}] = \sum_i [\textbf{C}]= 2 \sum_i [\textbf{D}]$ because of the extra reflection symmetry of $[\textbf{C}]$, which allows us to eliminate $[\textbf{C}]$ and $[\textbf{D}]$ from the equations in favor of $[\textbf{B}]$. The relations between these terms holds due to the summation over $i$ and marks the first appearance where global balance, rather than detailed balance, must be invoked.  Using the solutions at order $J$ and $J^2$, the global balance condition at order $J^3$ is,
% \begin{equation}
%     \sum_i \Big[(2  \tilde{j}_1^3 -2 j_1^3)[\textbf{A}] + (2\tilde{t}_4^3-2 t_4^3) [\textbf{B}] + 2\tilde{t}_4^3 [\textbf{C}]+ 2\tilde{t}_4^3 [\textbf{D}]+2\tilde{j}_2^3 [\textbf{H}]+ 2\tilde{j}_3^3 [\textbf{I}]\Big] =0.
% \end{equation}

\begin{equation}
    \sum_i \Big[(2  \tilde{j}_1^3 -2 j_1^3)[\textbf{A}] + (8\tilde{t}_4^3-2 t_4^3) [\textbf{B}]+2\tilde{j}_2^3 [\textbf{H}]+ 2\tilde{j}_3^3 [\textbf{I}]\Big] =0.
\end{equation}
Thus, we find $\tilde{j}_2^3=0$, $\tilde{j}_3^3=0$, and $\tilde{j}_1^3 = j_1^3$,  and $\tilde{t}_4^3 = t_4^3/4$.  At this order products of interactions also appear.  For example, one finds terms of the form $ [\textbf{A}]\cdot[\textbf{A}]\cdot[\textbf{A}]= 10[\textbf{A}]+6[\textbf{B}]$, however, all such terms have coefficients that vanish at this order.

Thus, to order $J^3$ there are only two terms in the sequential Hamiltonian,
 \begin{equation}
      \mathcal{-H} = \tilde{J}_1  \sum_{\langle i,j \rangle 1}S_i S_j + \tilde{T_4}  \sum_{\langle i,j,k,l \rangle \perp} S_i S_j S_k S_l +O[J^4],
 \end{equation}
where
\begin{align}
   \label{thirdorderJ1} \tilde{J}_1&=j_1^1 J + j_1^3 J^3  + O[J^4]&\\
    \nonumber &=\big[\tanh (K)+1\big] J +\frac{10}{3}  \big[\tanh ^3(K)-\tanh (K)\big] J^3 + O[J^4]&\\
    \nonumber \makebox{and} &\\
    \tilde{T}_4&=\frac{t_4^3}{4} J^3 + O[J^4]&\\
    \nonumber &=\frac{1}{2}\big[ \tanh ^3(K)- \tanh (K)\big] J^3 + O[J^4].
\end{align}
%\textcolor{red}{probably can eliminate next equation}
%Substituting back in the Eq. \eqref{eq3} we get,
%\begin{align}
% \nonumber \frac{P(\alpha(i))}{P(\alpha)} &=\exp\Big(-2 J_1 S_i^\alpha \sum_{\langle \rangle 1} S_j^\alpha - \frac{ T_4}{2} S_i^\alpha \sum_{\perp b} S_j^\alpha S_k^\alpha S_l^\alpha - \frac{ T_4}{2} S_i^\alpha \sum_{\perp c} S_j^\alpha S_k^\alpha S_l^\alpha- \frac{ T_4}{2} S_i^\alpha \sum_{\perp d} S_j^\alpha S_k^\alpha S_l^\alpha \Big)\\
% P(\alpha) &= \exp \Big( J_1  \sum_{\langle \rangle 1}S_i^\alpha S_j^\alpha + \frac{ T_4}{4}  \sum_{\perp} S_i^\alpha S_j^\alpha S_k^\alpha S_l^\alpha \Big)
%\end{align}
% and the Hamiltonian takes the form,
%  \begin{equation}
%       \mathcal{-H} = \tilde{J}_1  \sum_{\langle i,j \rangle 1}S_i S_j + \tilde{T_4}  \sum_{\langle i,j,k,l \rangle \perp} S_i S_j S_k S_l +O[J^4]
%  \end{equation}
% where $\sum_{\langle i,j \rangle 1}$ is sum over all the nearest neighbor interactions and $\sum_{\langle i,j,k,l \rangle \perp}$ is sum over all the interactions with four spins where three spins are in a line of length 2 units and the fourth spin is perpendicular to the center spin. 

We now consider additional couplings beyond those explicitly written in the Hamiltonian, Eq.\ \eqref{eq:Hansatz}. It is straightforward though tedious to show that no additional terms appear up to third order in $J$.  Consider an additional distinct coupling term in the Hamiltonian with coupling $\tilde{J}_x$. The ratio of the stationary probabilities before and after flipping spin $S_i$, Eq.\ \eqref{eq3}, will include a new set of terms, distinct from the previous terms, all rooted at site $i$, and all having coupling coefficients $\tilde{J}_x$.  As before, we expand $\tilde{J}_x$ in a power series in $J$, $\tilde{J}_x= \tilde{j}_x^1 J + \tilde{j}_x^2 J^2 + \tilde{j}_x^3 J^3 + \ldots$, and then supplement the global balance equation \eqref{eq:globalbalancebf} with the new terms.  Finally, one must carry out the expansion in powers of $J$ and see that the coefficients $\tilde{j}_x^1$ $\tilde{j}_x^2$ and  $\tilde{j}_x^3$ all vanish.  The intuitive reason why this is the case is that there are no dynamical couplings at order $J^3$ to balance an added term in the Hamiltonian.

On the other hand, at fourth and higher order in $J$, new types of interactions that are not present in the set shown in Fig.\ \ref{couplings} appear as a result of products of couplings such as  $[\textbf{A}]\cdot[\textbf{C}]$ and these interactions can couple spins not included  in the five spin ``cross''.    Both long range and multi-spin couplings are expected to appear at higher order in $J$ but as the range or number of spins increases, the leading order in $J$ of the associated coupling constant will increase.  The higher order terms in the sequential Hamiltonian are left for future research.

\subsubsection{Approximation to the sequential critical curve}
\label{sec:seqapprox}
Here we present an approximation to the critical dynamical coupling $J_c^{(s)}(K)$ for sequential dynamics.  Assume that the $\tilde{T}_4$ term and all higher order couplings can be neglected and that the nearest-neighbor coupling, $\tilde{J}_1$ is best approximated by the full expression $J_1(J,K)$ in Eq.\ \eqref{eq2} rather than the third order truncation, Eq.\ \eqref{thirdorderJ1}.  Then our approximation for the critical curve, $J_c^{(s)}(K)$ is given by the implicit equation,
\begin{equation}
\label{eq:seqJc}
 J_1(J_c^{(s)}(K),K)=J_c(0),  
\end{equation}
where $J_c(0) \approx 0.4407$ is the critial coupling of the ordinary Ising model.  The critical curve was obtained from this equation numerically and is shown in blue (upper curve) in Fig.\ \ref{fig:phase_plot}.

\begin{figure}
\centering
\includegraphics[width = .5\textwidth]{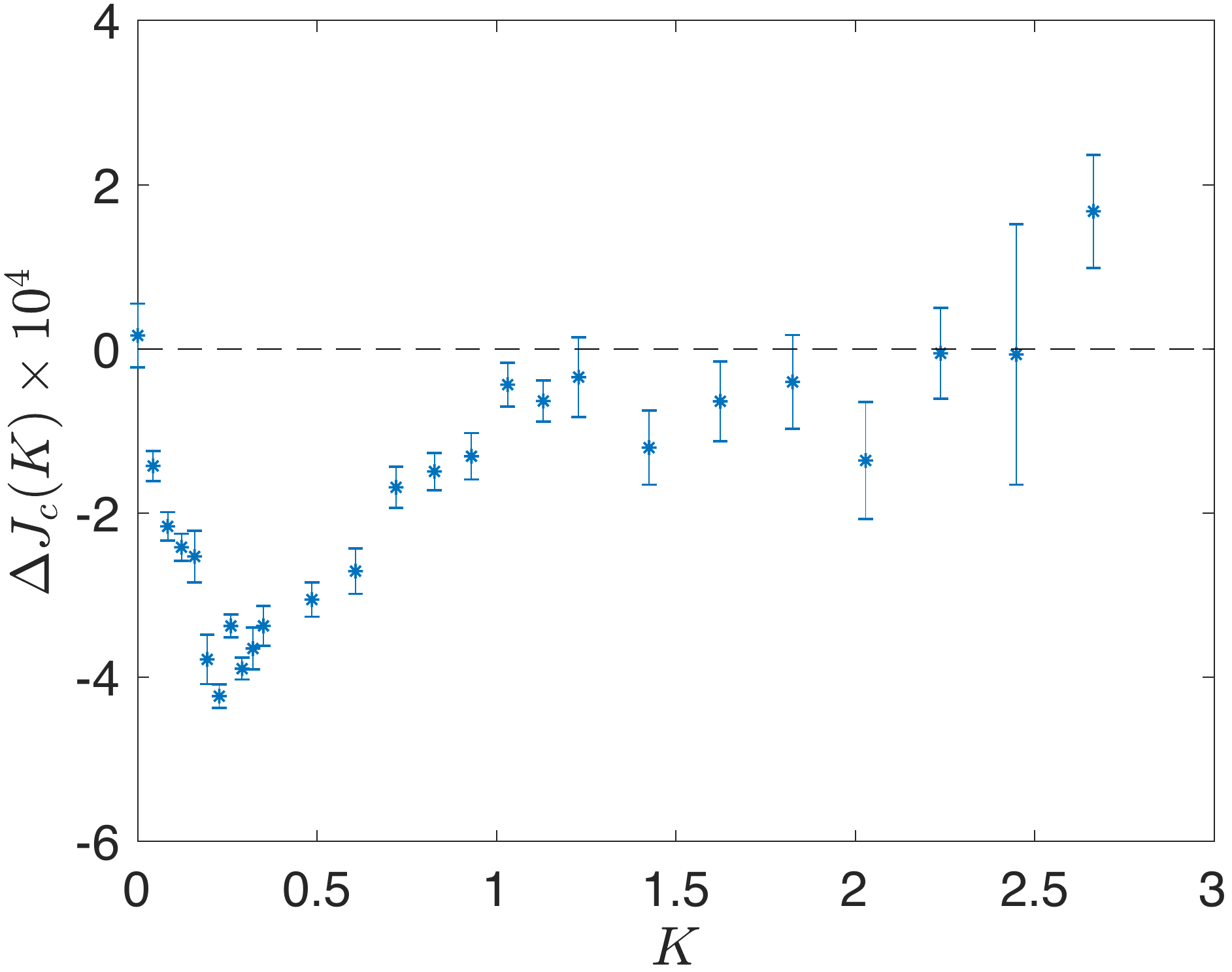}
\caption{\label{fig:difference} The difference, $\Delta J_{c}(K)$ between the numerical data for the critical curve for sequential dynamics and the approximation, Eq.\ \eqref{eq:seqJc}, as a function of $K$.} 
\end{figure}

This approximation is surprisingly good.  It is exact for both $K \rightarrow 0$ and $K \rightarrow \infty$.  Figure\ \ref{fig:difference} shows the difference, $\Delta J_c(K)$ between the numerical result for the critical coupling $J_c^{(s)}(K)$ and the approximation, $J_1(J_c^{(s)}(K),K)=J_c(0)$. Finite-size effects are not taken into account in  the numerical results and may be responsible for some of this difference, which never exceeds 4 parts in 10$^4$.  

% \begin{figure}
% \centering
% \includegraphics[width = .5\textwidth]{sequential_couplings.pdf}
% \caption{\label{fig:seq_couplings}The generated two and four spin couplings for sequential dynamics as defined in Eq.\ \eqref{eqn2} are plotted at $J = 1$ as a function of self-interaction $K$. The couplings satisfies the standard Ising model with twice the nearest neighbor interaction as $K \rightarrow \infty$. The system becomes  non-interactive as $K \rightarrow -\infty$.}
% \end{figure}

\section{Mean field theory}
\label{sec:mft}
In this section, the phase diagram for the kinetic Ising model with self-interaction %with dynamics given 
%by Eq.\ \eqref{eq:heatbath} 
is obtained within a mean field approximation. 
% The qualitative behavior of the mean field phase curve is compared with numerical results presented in Fig.\ \ref{fig:phase_plot}.
Mean field theory (MFT) for the equilibrium, nearest-neighbor Ising model with no self-interaction or magnetic field yields a self-consistent equation for the magnetization per spin $m$,
\begin{equation}
\label{eq:mfts}
m=\tanh(J zm)
\end{equation}
where $z$ is the coordination number. The critical coupling $J_c$ marking the onset of spontaneous magnetization ($|m|>0$)  occurs when $J_c z =1$.

One can interpret the mean field equation, \eqref{eq:mfts}, as a dynamical equation with the magnetization on the left hand side being one time step later than the magnetization on the right hand side. Equilibrium states are obtained by setting the magnetization on the two sides of the equation equal.
We can generalize the mean field dynamical equation to include self-interaction by replacing $h_i$ in Eq.\ \eqref{eq:heatbath} by $z m$. The mean field dynamical equation for spin, $S_i$ is,
\begin{equation}
    \langle S_i^\prime \rangle = \tanh(J zm + K S_i),
\end{equation}
where $\langle S_i^\prime \rangle$ is the expected value of spin $S_i$ at the subsequent time step.  We now average over sites $i$ to obtain the equilibrium equation for $m$,
\begin{equation}\label{mft}
m =\Big(\frac{1+m}{2}\Big) \tanh(J zm + K) + \Big(\frac{1-m}{2}\Big) \tanh(J zm - K),    
\end{equation}
where the first and second terms on the right hand side correspond to the the contributions from plus and minus spins, respectively.
% modify the mean field dynamics to include memory, consistent with Eq.\ \eqref{eq:heatbath} by including a memory term,
% \begin{equation}
% m =  \langle \tanh(J zm + K \sigma_i) \rangle
% \end{equation}
% where $\sigma_i$ is any of the $N$ spins and the average is over all $N$ spins. The average number of up spins, $n_+$ and down spins, $n_{-}$ can be written in terms of the magnetization, $n_\pm = (1 \pm m)/2$ and the self-consistent equation for $m$ becomes, 
% \begin{equation}\label{mft}
% n_+ =&\nonumber \frac{1+m}{2}; \qquad \qquad \quad \; \quad n_- = \frac{1-m}{2};\\
% m =&\Big(\frac{1+m}{2}\Big) \tanh(J zm + K) + \Big(\frac{1-m}{2}\Big) \tanh(J zm - K). 
% \end{align}

\begin{figure}
\centering
  \includegraphics[width = .5\textwidth]{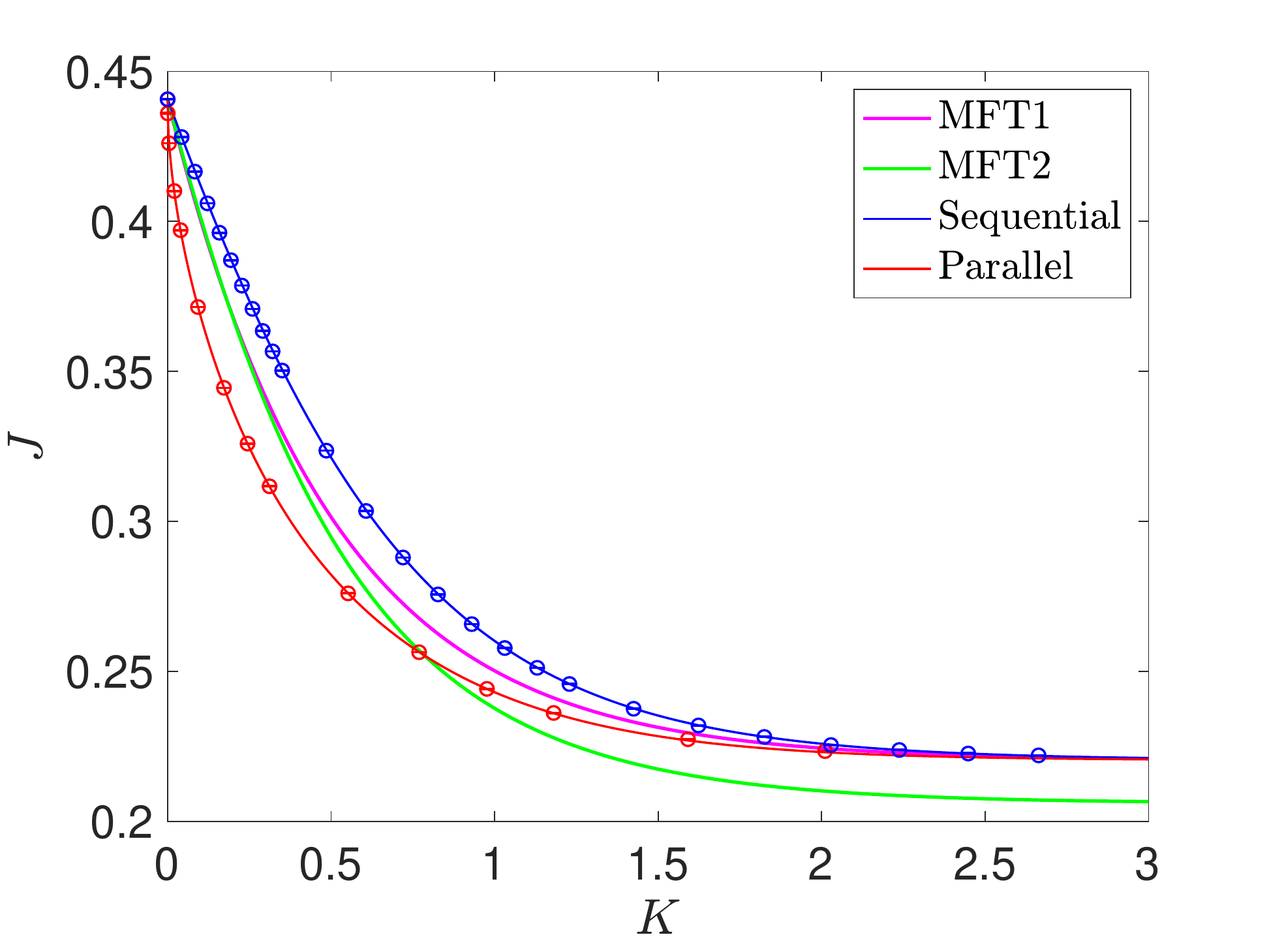}
\caption{\label{fig:mft} The phase diagram in the $J$-$K$ plane showing the critical dynamical coupling obtained from mean field theories. Critical lines for parallel (red, lower) and sequential dynamics (blue, upper) are shown for comparison(see Fig.\ \ref{fig:phase_plot} for details).  The magenta curve that lies between the data for sequential and parallel dynamics is the mean field theory, MFT1, obtained here (see Eq.\ \eqref{mft2}) and the green solid line that falls below the data points is mean field theory, MFT2, of Ref. \citep{CIRILLO201436}. }
\end{figure}

Close to the critical point, an expansion in $m$ yields an equation for the critical curve, $J_c(K)$,
\begin{equation}\label{mft2}
 J_c(K) z [ 1+\tanh(K)] - 1 = 0.
\end{equation}
 The phase diagram obtained from Eq.\ \eqref{mft2} is plotted as the solid magenta line in Fig.\ \ref{fig:mft}. The value of $z$ is adjusted so that $J_c(K=0)$ is the exact value for the two-dimensional Ising model. We see that MFT agrees qualitatively with the numerical results. Mean field theory does not distinguish between sequential and parallel dynamics so it is perhaps not surprising that the critical curve lies between the parallel and sequential critical curves.  Our mean field theory captures the exact result for both dynamics that $J_c(K) \rightarrow J_c(0)/2$ as $K \rightarrow \infty$. Finally, the mean field equation \eqref{mft2} is the same as the mean field equation for the standard Ising model except for the factor of $[1+\tanh(K)]$, which is precisely the first order in $J$ correction to the nearest-neighbor coupling in the sequential Hamiltonian (see Eq.\ \eqref{firstorder}).

An alternative mean field theory was obtained in \citep{CIRILLO201436} and is also shown in Fig.\ \ref{fig:mft} as the solid green curve. These results were calculated from a mean field approximation applied to the Hamiltonian obtained for parallel dynamics (see Eq.\ \eqref{eqn3}). The MFT phase plot for the parallel dynamics is adjusted so that $J_c(K=0)$ is the exact value for the two-dimensional Ising model. The qualitative behavior agrees with the other phase plots but it does not capture  $J_c(K) \rightarrow J_c(0)/2$ as $K \rightarrow \infty$.

\section{Discussions}
We have studied kinetic Ising models with nearest-neighbor dynamical interaction $J$ and self-interaction, $\mem$, and for both random sequential updating and parallel updating. These models were studied with Monte Carlo simulation and analytic methods.  The equilibrium phase diagram and critical lines of both models were obtained numerically and approximated by several methods.  One of these approximations is a simple mean field theory that predicts a critical line that falls between the parallel and sequential critical lines. In addition, the critical dynamics of the sequential model was studied and the prefactor of the critical divergence of the magnetization autocorrelation time was found to increase as $e^{2 \mem}$. 

We have studied the Gibbs distributions describing the equilibrium states of the two models.  For the parallel model, this distribution satisfies detailed balance with respect to the dynamics. The associated Hamiltonian was previously obtained \citep{CIRILLO201436}, and involves three two-spin couplings and two four-spin couplings.  The equilibrium distribution for the random sequential model does not satisfy detailed balance with respect to the dynamics and must be obtained from the global balance equation.  We developed a perturbative expansion of the sequential Hamiltonian in the dynamical coupling, $J$ and carried it out to order $J^3$.  At this order the Hamiltonian features a nearest-neighbor coupling and a four-spin coupling. It is clear that at higher order, other smaller couplings, perhaps infinitely many other couplings, will appear.  The nearest-neighbor coupling obtained from the perturbative expansion yields a very accurate approximation to the critical line.  It would be interesting to understand the equilibrium Gibbs distribution for sequential updating more completely.  Does the Hamiltonian in fact have infinitely many coupling and, if so, how do they fall off as a function of the range of interaction and the number of coupled spins? 

For the parallel case we focused attention on the region near the $\mem=0$ critical point, which is described by two uncoupled critical Ising models on the two sublattices of the square lattice. When the full lattice is viewed, large checkerboard regions are apparent at and near this critical point. For small, but non-vanishing $\mem$ two critical lines emerge from the $\mem=0$ critical point for positive and negative $\mem$, respectively.  Numerical results show that these critical lines approach the $\mem=0$ critical point as a power law with an exponent slightly larger than $1/2$.  We proposed a finite-size scaling theory for the region near the $\mem=0$ critical point that includes a new critical exponent that controls the scaling behavior in the $\mem$ direction and the power-law behavior of the critical lines.  We presented a heuristic argument based on the length scale for checkerboard regions as a function of $\mem$, which predicts that the exponent describing the shape of the critical lines is $1/2$.  It would be interesting to understand the scaling behavior near the $\mem=0$ critical point more fully.  Is the finite-size scaling theory correct? If so, what is the actual scaling exponent and can it be related to known Ising critical exponents.  

When self-interaction is large and positive, the equilibrium states of the sequential and parallel models both reduce to a nearest-neighbor Ising model with twice the dynamical coupling strength \citep{Da2012}. The equivalence of the two updating schemes can be understood intuitively from the fact that spin-flips occur exponentially rarely so that a parallel update becomes equivalent to a single sweep of sequential updates--in both cases almost no spin flips occur.

\acknowledgements
We thank Karen Abbott, Shadi Esmaelli, Alan Hastings,  and Andrew Noble for useful discussions.  The work was supported in part by NSF grant DMS-1840221.

\renewcommand\refname{Bibliography}

\bibliographystyle{unsrt}
\bibliography{references}

\end{document}